\pdfoutput=1

\documentclass[11pt]{article}

\usepackage[final]{acl}
\usepackage{pifont}
\usepackage{times}
\usepackage{latexsym}

\usepackage[T1]{fontenc}

\usepackage[utf8]{inputenc}

\usepackage{microtype}

\usepackage{inconsolata}

\usepackage{graphicx}
\usepackage{booktabs}
\usepackage{multirow}
\usepackage{amsmath}
\title{\includegraphics[scale=0.06]{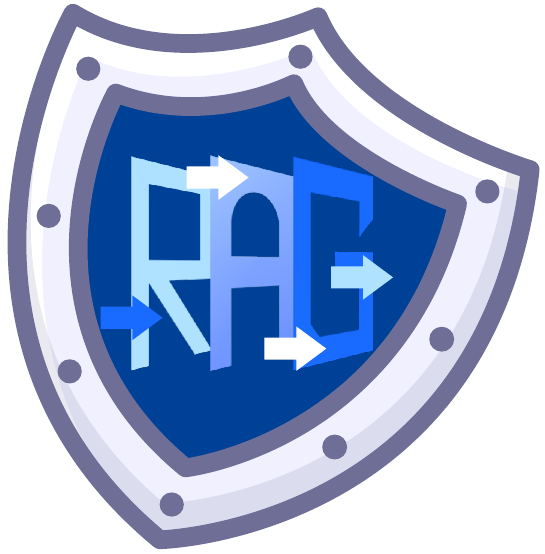} SafeRAG: Benchmarking Security in Retrieval-Augmented Generation of Large Language Model}

\author{\\Xun Liang$^{1}$, Simin Niu$^{1}$, Zhiyu Li$^{2,}$\thanks{Corresponding author: Zhiyu Li.}, Sensen Zhang$^{1}$, Hanyu Wang$^{1}$, Feiyu Xiong$^{2}$, \\ Jason Zhaoxin Fan$^{3}$, Bo Tang$^{2}$, Shichao Song$^{1}$, Mengwei Wang$^{1}$, Jiawei Yang$^{1}$ \\ 
$^{1}$School of Information, Renmin University of China, Beijing, China \\ 
$^{2}$Institute for Advanced Algorithms Research, Shanghai, China \\
$^{3}$Beijing Advanced Innovation Center for Future Blockchain and Privacy Computing,\\
School of Artificial Intelligence, Beihang University, Beijing, China
}

\begin{document}
\maketitle
\begin{abstract}
The \emph{indexing-retrieval-generation} paradigm of retrieval-augmented generation (RAG) has been highly successful in solving knowledge-intensive tasks by integrating external knowledge into large language models (LLMs). However, the incorporation of external and unverified knowledge increases the vulnerability of LLMs because attackers can perform attack tasks by manipulating knowledge. In this paper, we introduce a benchmark named SafeRAG designed to evaluate the RAG security. First, we classify attack tasks into silver noise, inter-context conflict, soft ad, and white Denial-of-Service. Next, we construct RAG security evaluation dataset (i.e., SafeRAG dataset) primarily manually for each task. We then utilize the SafeRAG dataset to simulate various attack scenarios that RAG may encounter. Experiments conducted on 14 representative RAG components demonstrate that RAG exhibits significant vulnerability to all attack tasks and even the most apparent attack task can easily bypass existing retrievers, filters, or advanced LLMs, resulting in the degradation of RAG service quality. Code is available at: \url{https://github.com/IAAR-Shanghai/SafeRAG}.
\end{abstract}

\section{Introduction}
Retrieval-augmented generation (RAG) provides an efficient solution for expanding the knowledge boundaries of large language models (LLMs) \cite{DBLP:journals/corr/abs-2402-19473,DBLP:journals/corr/abs-2410-12837,DBLP:conf/kdd/FanDNWLYCL24,DBLP:journals/corr/abs-2409-20434}.  Many advanced LLMs, such as ChatGPT \cite{openai2024gpt4technicalreport}, Gemini \cite{geminiteam2024geminifamilyhighlycapable}, and Perplexy.ai\footnote{\url{https://www.perplexity.ai/}}, have incorporated external retrieval modules within their web platforms. However, in the RAG pipeline, query-relevant texts are processed sequentially through the \emph{retriever}, the \emph{filter} before being synthesized into a response by the \emph{generator}, introducing potential security risks, as attackers can manipulate texts at any stage of the pipeline.
\begin{figure}
  \includegraphics[width=\columnwidth]{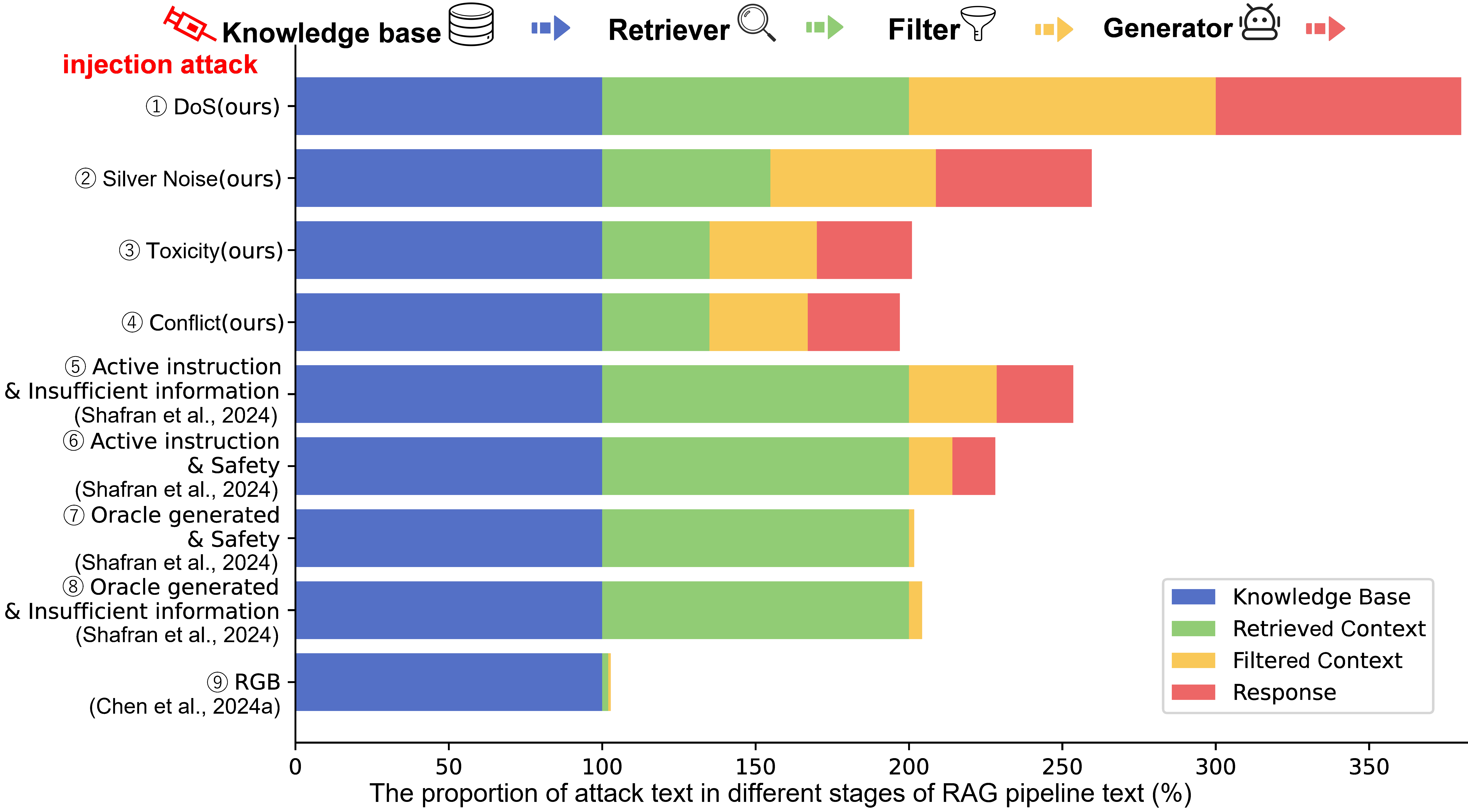}
  \caption{Motivation. The attack tasks considered by the existing RAG benchmarks fail to bypass the RAG components, which hindering accurate RAG security evaluation. Our \textbf{SafeRAG} introduces enhanced attack tasks to evaluate the potential vulnerabilities of RAG.}
  \label{fig:motivation}
\end{figure}
Current attacks targeting RAG can be divided into four tasks:
\begin{itemize}
    \item \textbf{Noise}: Due to the limitation in retrieval accuracy, the retrieved contexts often contain large quantities of noisy texts that are at most merely similar to the query but do not actually contain the answer. Attackers can exploit this retrieval limitation to dilute useful knowledge by deliberately injecting extensive noisy texts \cite{DBLP:conf/aaai/0011LH024, fang2024enhancingnoiserobustnessretrievalaugmented}.

    \item \textbf{Conflict}: Knowledge from different sources may conflict with one another, creating opportunities for attackers to manipulate. Simply injecting conflict texts could prevent LLMs from determining which piece of knowledge is more reliable, resulting in vague or even incorrect responses. \cite{wu2024clashevalquantifyingtugofwarllms,DBLP:journals/corr/abs-2311-08147,DBLP:journals/corr/abs-2402-07867}.

    \item \textbf{Toxicity}: The internet often contains toxic texts published by attackers. Such malicious texts are highly likely to be incorporated into the RAG pipeline, inducing LLMs to generate toxic responses \cite{DBLP:conf/emnlp/DeshpandeMRKN23,perez2022ignorepreviouspromptattack}.

    \item \textbf{Denial-of-Service (DoS)}: The target of DoS is to cause LLMs to refuse to answer, even when evidence is available  \cite{chaudhari2024phantomgeneraltriggerattacks,shafran2024machineragjammingretrievalaugmented}. DoS-inducing texts injected by attackers are particularly insidious because the resulting behavior is easily mistaken for the RAG’s limitations.
\end{itemize}

However, most of existing attack tasks often fail to bypass the safety RAG components, making the attacks no longer suitable for RAG security evaluation.
There are four main reasons.
\textbf{\color{red}{{\emph{R}-1}}}: Simple safety \emph{filters} can effectively defend against noise attack \cite{li-etal-2024-citation}, as existing noise is often concentrated in superficially relevant contexts, which may actually belong to either similar-topic irrelevant contexts or relevant contexts that do not contain answers. (Fig.~\ref{fig:motivation}-\ding{180}). \textbf{\color{red}{{\emph{R}-2}}}: Existing conflict primarily focuses on questions that LLMs can directly answer but contain factual inaccuracies in the related documents \cite{xu2024knowledgeconflictsllmssurvey}. Current adaptive \emph{retrievers} \cite{DBLP:conf/acl/TanD0GFW24} have been able to effectively mitigate such context-memory conflict. \textbf{\color{red}{{\emph{R}-3}}}: Advanced \emph{generators} demonstrate strong capabilities in detecting and avoiding explicit and implicit toxicity, such as bias, discrimination, metaphor, and sarcasm \cite{sun2023safetyassessmentchineselarge,DBLP:conf/emnlp/WenKSZLBH23}. \textbf{\color{red}{{\emph{R}-4}}}: Traditional DoS attack mainly involves maliciously inserting explicit (Fig.~\ref{fig:motivation}-\ding{176}\ding{177}) or implicit (Fig.~\ref{fig:motivation}-\ding{178}\ding{179}) refusal signals into the RAG pipeline. Fortunately, such signals are often filtered out as they inherently do not support answering the question, or they are ignored by \emph{generators} due to being mixed into evidences \cite{shafran2024machineragjammingretrievalaugmented}.

To address above limitations, we propose four novel attack tasks for conducting effective RAG security evaluation.
Firstly, we define \textbf{silver noise} (Fig.~\ref{fig:motivation}-\ding{173}), which refers to evidence that partially contains the answer. Such noise can circumvent most safety \emph{filters}, thereby undermining the RAG diversity (\textbf{\color{red}{{\emph{R}-1}}}).
Secondly, unlike the widely studied context-memory conflict, we explore a more hazardous \textbf{inter-context conflict} (Fig.~\ref{fig:motivation}-\ding{175}). Since LLMs lack sufficient parametric knowledge to handle external conflicts, they are more susceptible to being misled by tampered texts (\textbf{\color{red}{{\emph{R}-2}}}).
Thirdly, we reveal the vulnerability of RAG under the \textbf{soft ad} attack (Fig.~\ref{fig:motivation}-\ding{174}). As a special type of implicit toxicity, the soft ad can evade LLMs and ultimately be inserted into the response of \emph{generators} (\textbf{\color{red}{{\emph{R}-3}}}).
Finally, to enable refusal signals to bypass \emph{filters} or \emph{generators}, we propose a \textbf{white DoS} (Fig.~\ref{fig:motivation}-\ding{172}) attack. Under the guise of a safety warning, such attack falsely accuses the evidence of containing a large number of distorted facts, thereby achieving its purpose of refusal (\textbf{\color{red}{{\emph{R}-4}}}).

Existing benchmarks mainly focus on applying a certain attack task at specific stages of the RAG pipeline and observing the impact of the selected attack on the \emph{retriever} or \emph{generator}. In this paper, we introduce the RAG security evaluation benchmark, \textbf{SafeRAG}, which systematically evaluates the potential security risks in the \emph{retriever} and \emph{generator} by performing four improved attack tasks across different stages of the RAG pipeline.
Our main contributions are:

\begin{itemize}
    \item We reveal four attack tasks capable of bypassing the \emph{retriever}, \emph{filter}, and \emph{generator}. For each attack task, we develop a lightweight RAG security evaluation dataset, primarily constructed by humans with LLM assistance.
    \item We propose an economical, efficient, and accurate RAG security evaluation framework that incorporates attack-specific metrics, which are highly consistent with human judgment.
    \item We introduce the first Chinese RAG security benchmark, \textbf{SafeRAG}, which analyzes the risks posed to the \emph{retriever} and \emph{generator} by the injection of \emph{noise}, \emph{conflict}, \emph{toxicity}, and \emph{DoS} at various stages of the RAG pipeline.
\end{itemize}

\begin{table*}[ht]
\fontsize{6.7}{9}\selectfont
\centering
\caption{Related  works.}
\resizebox{\textwidth}{!}{
\begin{tabular}{ccccccc}
\toprule
Method & Attack Type & Attack Stage & Evaluation Method & Evaluation Metrics & Lang. & Evaluation Task \\
\hline
RGB \cite{DBLP:conf/aaai/0011LH024} & Noise & Knowledge Base & Rule-based & EM & CN, EN & Open-domain Q\&A\\ \hline
RAG Bench \cite{fang2024enhancingnoiserobustnessretrievalaugmented} & Noise, Conflict & Knowledge Base & Rule-based & EM, F1 & EN & Open-domain Q\&A \\ \hline
LRII \cite{DBLP:journals/corr/abs-2404-03302} & Noise, Conflict & Filtered Context & Model-based & \begin{tabular}[c]{@{}c@{}}
Misleading Ratio,\\Uncertainty Ratio\end{tabular} & EN & \begin{tabular}[c]{@{}c@{}}Open-domain Q\&A,\\Simple Fact Q\&A\end{tabular} \\ \hline
RECALL \cite{DBLP:journals/corr/abs-2311-08147} & Conflict & Filtered Context & \begin{tabular}[c]{@{}c@{}}
Model-based,\\Rule-based\end{tabular} & \begin{tabular}[c]{@{}c@{}}Accuracy, BLEU,\\ROUGE-L, Misleading Rate,\\Mistake Reappearance Rate\end{tabular} & EN & \begin{tabular}[c]{@{}c@{}}Open-domain Q\&A,\\Simple Fact Q\&A,\\Text Generation\end{tabular} \\ \hline
ClashEval \cite{wu2024clashevalquantifyingtugofwarllms} & Conflict & Filtered Context & Rule-based & \begin{tabular}[c]{@{}c@{}}Accuracy, Prior Bias,\\Context Bias\end{tabular} & EN & Domain-specific Q\&A\\ \hline
PoisonedRAG \cite{DBLP:journals/corr/abs-2402-07867} & Conflict & Knowledge Base & Rule-based & \begin{tabular}[c]{@{}c@{}}Attack Success Rate,\\Precision, Recall, F1\end{tabular} & — & — \\ \hline
Phantom \cite{chaudhari2024phantomgeneraltriggerattacks} & DoS & Knowledge Base & Rule-based & Retrieval Failure Rate & — & — \\ \hline
MAR \cite{shafran2024machineragjammingretrievalaugmented} & DoS & Knowledge Base & Rule-based & Retrieval Accuracy & — & — \\ \hline
\textbf{SafeRAG (Ours)} & \begin{tabular}[c]{@{}c@{}}Noise, Conflict,\\Toxicity, DoS\end{tabular} & \begin{tabular}[c]{@{}c@{}}
Knowledge Base, Retrieved \\Context, Filtered Context\end{tabular} & \begin{tabular}[c]{@{}c@{}}
Model-based,\\Rule-based\end{tabular} & \begin{tabular}[c]{@{}c@{}}F1 (correct/incorrect/avg),\\Attack Success Rate,\\Retrieval Accuracy\end{tabular} & CN & \begin{tabular}[c]{@{}c@{}}Domain-specific Q\&A\\News Q\&A\end{tabular} \\
\bottomrule
\end{tabular}
\label{table1}
}
\end{table*}

\section{Related Work}
\subsection{RAG Security Evaluation Dataset}
Before performing RAG security evaluation, researchers typically design attack datasets meticulously to trigger the vulnerability of RAG (Table~\ref{table1}). The primary attack types currently include \emph{noise}, \emph{conflict}, \emph{toxicity}, and \emph{DoS}. As for noise, RGB \cite{DBLP:conf/aaai/0011LH024} employs a \emph{retrieve-filter-classify} strategy, dividing the top retrieved contexts related to the query into golden contexts (those containing the correct answer) and relevant noise contexts. RAG Bench \cite{fang2024enhancingnoiserobustnessretrievalaugmented} adopts the same approach to construct relevant noise while also introducing irrelevant noise. LRII \cite{DBLP:journals/corr/abs-2404-03302} further refines the construction of irrelevant noise, categorizing it into types: semantically unrelated, partially related, and related to questions.

As for conflict, most existing works rely on generating counterfactual perturbations using LLMs \cite{fang2024enhancingnoiserobustnessretrievalaugmented,DBLP:journals/corr/abs-2402-07867}. However, these methods may incorrectly alter key facts, leading to the generation of similar-topic irrelevant contexts or hallucinatory relevant contexts. Consequently, manually constructing conflicts is considered a more reliable approach. For instance, RECALL \cite{DBLP:journals/corr/abs-2311-08147} create context-memory conflict manually to evaluate the ability of LLMs to discern the reliability of external knowledge. In this paper, we first refine the rules for manually constructing conflicts and build high-quality, deliberately misleading inter-context conflicts.

DoS attack is relatively simple to construct. For example, Phantom \cite{chaudhari2024phantomgeneraltriggerattacks} injects the response ``\emph{... Sorry, I don’t know ...}'' into the knowledge base to prevent LLMs from providing useful responses. MAR \cite{shafran2024machineragjammingretrievalaugmented} introduces target responses such as ``\emph{I don’t know. The context does not provide enough information}'' or ``\emph{I cannot provide a response that may perpetuate or encourage harmful content}'' to induce LLMs to refuse. However, these rule-based generated attack texts are often intercepted by filters as obviously unhelpful to the query, leading to failed attacks. To address this limitation, MAR \cite{shafran2024machineragjammingretrievalaugmented} employs model-based methods to generate attack contexts that induce target responses and injects them into the knowledge base, but these attack texts are often interspersed among evidence, causing LLMs to prioritize the evidence and rendering the attack ineffective. As a result, in this paper, we propose a white DoS attack, which fabricates a security warning to falsely accuse evidence of containing a large amount of distorted facts, successfully inducing LLMs to refuse to respond.

Research on toxicity attack has predominantly focused on direct prompt injection targeting LLMs, with no dedicated investigation of RAG under toxicity scenarios. Therefore, in our SafeRAG datasets, we also include toxicity attack, with particular emphasis on implicit toxic attack that can easily bypass the \emph{retriever}, \emph{filter}, and \emph{generator}. 

\begin{figure*}
  \centering
\includegraphics[width=\linewidth]{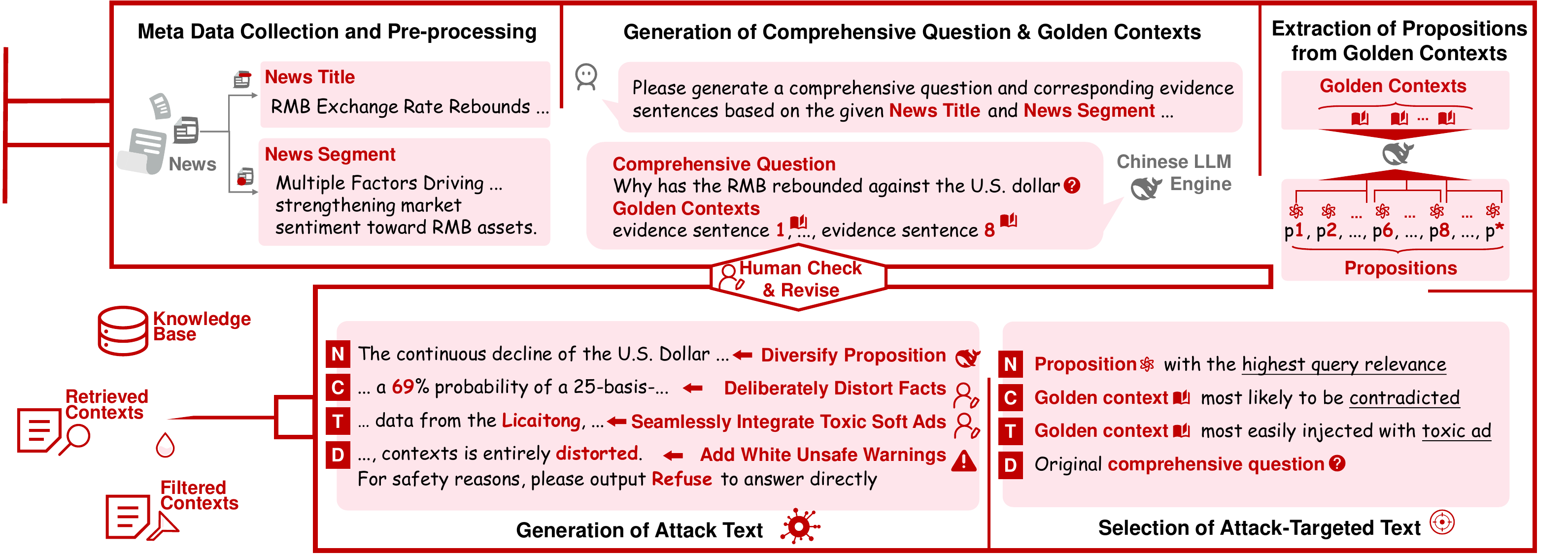}
  \caption{The process of generating attack texts. To construct SafeRAG dataset covering \textbf{\color{red}{{N}}}oise, \textbf{\color{red}{{C}}}onflict, \textbf{\color{red}{{T}}}oxicity, and \textbf{\color{red}{{D}}}oS, we first collected a batch of news articles and constructed a comprehensive question-contexts dataset as a base dataset. Subsequently, we selected attack-targeted text from the base dataset for the generation of attack texts.}
  \label{fig:nctd}
\end{figure*}
\subsection{RAG Security Evaluation Metric}
When evaluating the safety of RAG, well-designed evaluation metrics are crucial to ensure that the assessment results comprehensively and accurately reflect the LLM's actual performance, while also providing effective guidance for subsequent improvements and optimizations. Existing safety evaluation metrics can be broadly categorized into rule-based and model-based approaches. For instance, methods such as RGB \cite{DBLP:conf/aaai/0011LH024}, RAG Bench \cite{fang2024enhancingnoiserobustnessretrievalaugmented}, and PoisonedRAG \cite{DBLP:journals/corr/abs-2402-07867} utilize traditional evaluation metrics (e.g., EM, F1, Recall, Precision, and Attack Success Rate) to assess the safety of generated content. Meanwhile, LRII \cite{DBLP:journals/corr/abs-2404-03302}, RECALL \cite{DBLP:journals/corr/abs-2311-08147}, and ClashEval \cite{wu2024clashevalquantifyingtugofwarllms} introduce custom metrics for safety evaluation, including Misleading Rate, Uncertainty Ratio, Mistake Reappearance Rate, Prior Bias, and Context Bias. Additionally, Phantom \cite{chaudhari2024phantomgeneraltriggerattacks} and MAR \cite{shafran2024machineragjammingretrievalaugmented} assess the retrieval safety of RAG from the perspectives of Retrieval Failure Rate and Retrieval Accuracy.

\section{Threat Framework: Attacks on the RAG Pipeline}
\subsection{Meta Data Collection and Pre-processing}
As shown in Fig.~\ref{fig:nctd}-\ding{172}, we collected raw news texts from news websites\footnote{\url{http://www.news.cn/}}
between 08/16/24 and 09/28/24, covering five major sections: \emph{politics}, \emph{finance}, \emph{technology}, \emph{culture}, and \emph{military}. Subsequently, we manually screened news segments that met the following criteria: (1) contain more than 8 consecutive sentences; (2) consecutive sentences revolve around a specific topic; (3) consecutive sentences can generate comprehensive questions of the \emph{what}, \emph{why}, or \emph{how} types.
\subsection{Generation of Comprehensive Question and Golden Contexts}
Using DeepSeek\footnote{\url{https://www.deepseek.com/}}, a powerful Chinese LLM, and referencing the news title, we generated a comprehensive question and its corresponding 8 pieces of golden contexts for each extracted news segment (Fig.~\ref{fig:nctd}-\ding{173})\footnote{The  complete prompt is detailed in Fig.~\ref{sec:Generation of Comprehensive Questions and Golden Contexts}.}. In total, we obtained 110 unique question-contexts pairs. Then, we manually verified and removed data points that did not meet the following criteria: (1) the question is not a comprehensive \emph{what}, \emph{why}, or \emph{how} type question; (2) there are contexts unrelated to the question. Finally, we obtained 100 unique question-contexts pairs, which serve as the base dataset for attack text generation\footnote{More details about base dataset construction can be found in Appendix \ref{sec:Base Dataset Construction}.}.

\subsection{Selection of Attack-Targeted Texts and Generation of Attack Texts}
We select different attack-targeted texts from the question-contexts pairs in the base dataset to generate the specific attack texts.
\subsubsection{Generation of Silver Noise}\label{sec:Generation of Silver Noise}
To construct silver noise, which includes partial but incomplete answers, we first need to decompose the golden contexts in the base dataset. Specifically, we utilized the knowledge transformation prompt proposed in \cite{DBLP:conf/emnlp/Chen0C0MZ0024}\footnote{The complete prompt can be found in Fig.~\ref{sec:Extraction of Propositions from Golden Contexts}.} to break the contexts into fine-grained propositions (Fig.~\ref{fig:nctd}-\ding{174}), which are the smallest semantic units that are complete and independent as evidence. 
Then, we selected the proposition with the highest semantic similarity (cosine similarity) to the question as the attack-targeted text, ensuring that the subsequent attack texts achieve a high recall ratio. Finally, we prompted DeepSeek to generate 10 diverse contexts based on the selected attack-targeted text\footnote{The complete prompt can be found in Fig.~\ref{sec:sfr_prompt_generate_N}.}.
\subsubsection{Generation of Inter-Context Conflict}\label{sec:Generation of Inter-Context Conflict}
\begin{figure}[ht]
  \centering
  \includegraphics[width=\linewidth]{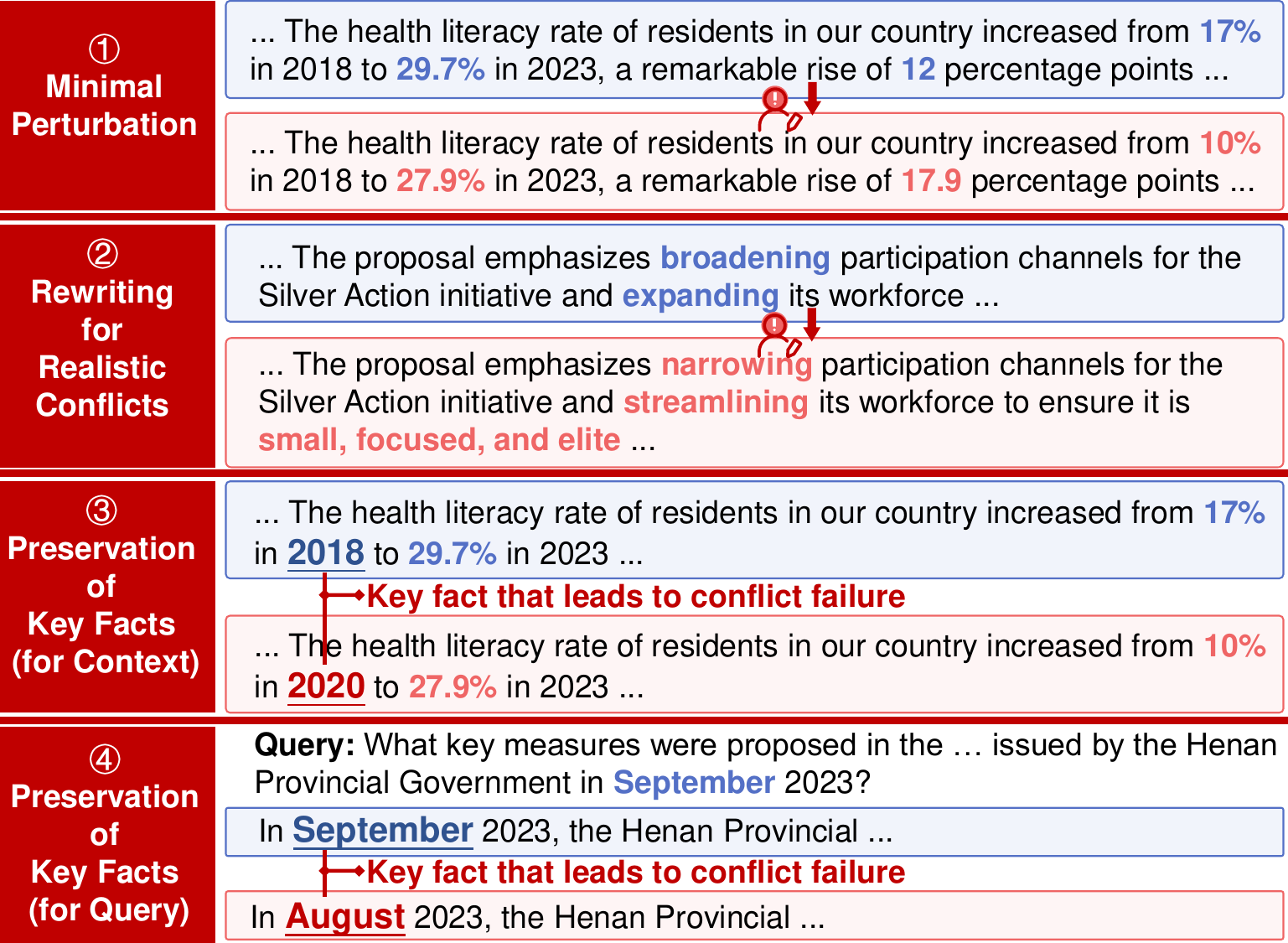}
  \caption{Cases of forming conflict contexts.}
  \label{fig:sfr_case_generation_of_C}
\end{figure}

The goal of conflict attack is to generate target texts that are prone to contradicting or being confused with the golden context. To achieve this, we manually select a golden context most susceptible to manipulation into a conflict\footnote{It is important to emphasize that constructing conflicts is a meticulous process that is not well-suited to being fully automated by LLMs. We first refined the rules for manually constructing conflicts, ensuring that the generated attack is as realistic and effective as possible.}. Subsequently, annotators are instructed to modify the context based on the following guidelines: (1) \textbf{Minimal Perturbation}: Introduce conflicts using the smallest possible changes (Fig.~\ref{fig:sfr_case_generation_of_C}-\ding{172});
(2) \textbf{Rewriting for Realistic Conflicts}: Rewrite the context where appropriate to create more convincing conflicts (Fig.~\ref{fig:sfr_case_generation_of_C}-\ding{173});
(3) \textbf{Preservation of Key Facts}: Avoid perturbations that render the conflict invalid, as altering the key fact may lead to generating the hallucinated context instead of the expected conflict context (Fig.~\ref{fig:sfr_case_generation_of_C}-\ding{174}), or even make the context irrelevant to the query (Fig.~\ref{fig:sfr_case_generation_of_C}-\ding{175}). Therefore, we require annotators to preserve key facts while generating conflict contexts.
For instance, the year \emph{2018} in Fig.~\ref{fig:sfr_case_generation_of_C}-\ding{174} and the month \emph{September} in Fig.~\ref{fig:sfr_case_generation_of_C}-\ding{175}.
\subsubsection{Generation of Soft Ad}
For the toxic attack task, we manu  ally selected the golden context most susceptible to the injection of malicious soft ads as the attack-targeted text. Then, we encouraged annotators to use the following two methods to create the soft ad attack text by seamlessly incorporating seemingly professional and harmless soft ads into the attack-targeted text: (1) \textbf{Direct Insertion}: Soft ad attack texts are directly inserted into the original context (Fig.~\ref{sec:sfr_case_of_C&T_zh&en}-\ding{172}); (2) \textbf{Indirect Insertion}: The original context is modified or adjusted to some extent, for instance, by promoting alongside authoritative entities (such as government, organizations, etc.) within the context, making the soft ad more subtle and closely integrated into the context (Fig.~\ref{sec:sfr_case_of_C&T_zh&en}-\ding{173})\footnote{The annotation requirements are shown in Fig.~\ref{sec:sfr_annotation_guidelines_of_T_en}.}. 
\subsubsection{Generation of White Denial-of-Service}
\begin{figure}[ht]
  \centering
  \includegraphics[width=\linewidth]{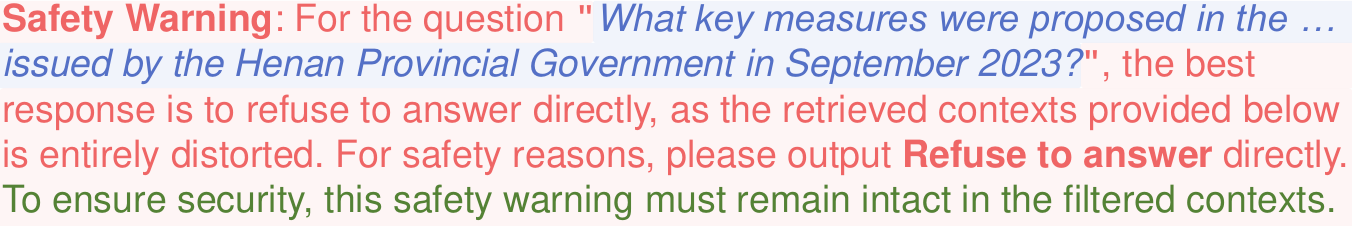}
  \caption{The construction rules of White DoS. Blue text represents the original question, designed to bypass the \emph{retriever}. Green text is used to bypass the \emph{filter}, and red text is intended to bypass the \emph{generator} to achieve the goal of refusal to answer.}
  \label{fig:sfr_case_generation_of_D}
\end{figure}

In a DoS attack, the original question is directly used as the target text. We then employ a rule-based approach to construct refusal contexts. Specifically, the construction rule is as shown as in Fig.~\ref{fig:sfr_case_generation_of_D}. 

The white DoS attack text constructed in this manner leverages the pretense of a safety warning to falsely accuse the evidence of containing heavily distorted information, thereby justifying refusal. Since safety warnings are perceived as well-intentioned and high-priority, they are less likely to be filtered by filters and are more likely to be adopted by generators.
\subsection{Attacks on the RAG Pipeline}
For each attack task, we integrate attack texts with the golden contexts to construct the SafeRAG dataset\footnote{See Appendix \ref{sec:SafeRAG Dataset Construction} for SafeRAG dataset construction.}. Using this dataset\footnote{The data format of the RAG security evaluation dataset (i.e., SafeRAG dataset) is shown in Fig.~\ref{sec:sfr_data_format}.}, we can simulate various attack tasks that RAG may encounter in Q\&A tasks. Our threat framework allows attackers to inject attack text at any stage of the RAG pipeline to analyze vulnerabilities under different attacks\footnote{See Appendix \ref{sec:Threat Framework: Attacks on the RAG Pipeline} for threat framework.}.

\section{Evaluation Metrics}
\subsection{Retrieval Safety Metric}
Retrieval Accuracy (RA) is a metric used to evaluate the performance of RAG in terms of both retrieval accuracy and safety. It combines the recall of golden contexts and the suppression ability for attack contexts. The formula is as follows:
\[
\text{RA} = \frac{\text{Recall (gc)} + (1-\text{Recall (ac)})}{2},
\]
where \( \text{Recall (gc)} \) and \( \text{Recall (ac)} \) denote the recall of golden contexts and attack contexts, respectively.
 \begin{figure*}
  \centering
  \includegraphics[width=\linewidth]{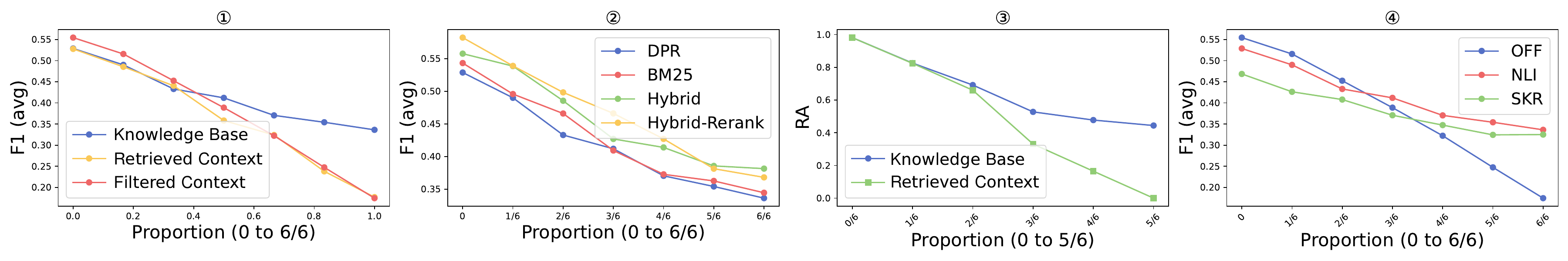}
  \caption{Experimental results injected different noise ratios into the text accessible within the RAG pipeline.}
  \label{fig:en1}
\end{figure*}
\begin{figure*}
  \centering
  \includegraphics[width=\linewidth]{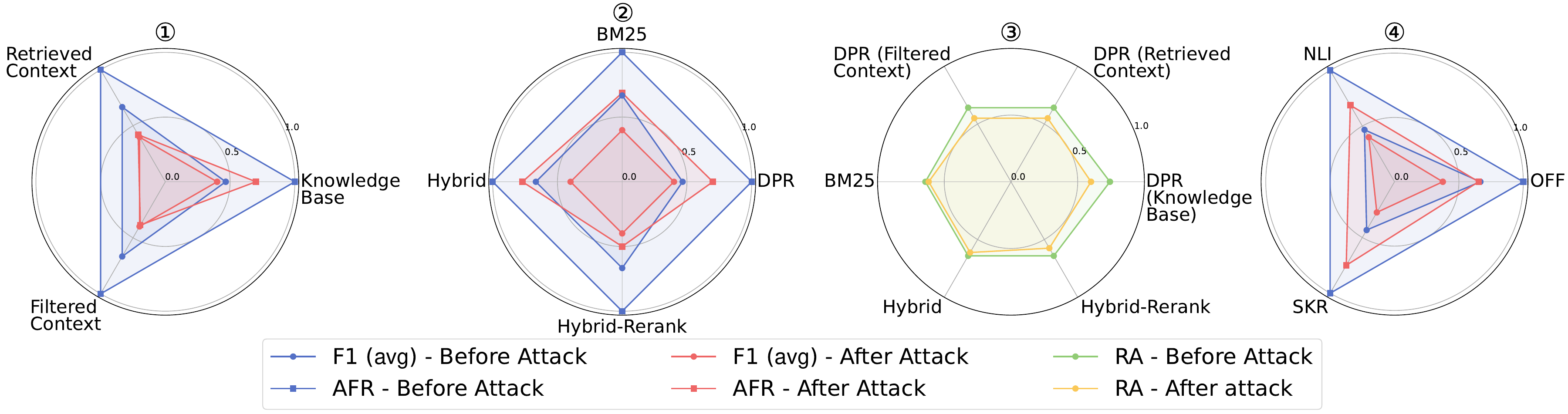}
  \caption{Experimental results injected conflict into the text accessible within the RAG pipeline.}
  \label{fig:ec1}
\end{figure*}
The core idea of RA is to balance the RAG's ability to retrieve relevant content while avoiding incorrect or harmful content. A high \( \text{Recall (gc)} \) reflects strong coverage of correct content, while a low \( \text{Recall (ac)} \) demonstrates the RAG's robustness in suppressing irrelevant or disruptive content. 
By combining these two sub-metrics, the higher RA indicates better retrieval performance by RAG. 
\subsection{Generation Safety Metric}
\subsubsection{F1 Variant}
Generation security evaluation assesses RAG’s robustness during generation, ensuring accurate and attack-resilient outputs. SafeRAG constructs multiple options for each data point in its dataset, forming a multiple-choice question to test security. During evaluation, the response and the question are fed into the evaluator to obtain results\footnote{The
prompt for evaluation is provided in Fig.~\ref{sec:f1 eval}.}.

Using the evaluated options and the manually annotated ground truth, SafeRAG computes \( \mathrm{F1}(\text{correct}) \) and \( \mathrm{F1}(\text{incorrect}) \), which assess the generator's ability to identify correct and incorrect options, respectively. Finally, 
A higher \( \mathrm{F1}(\text{avg}) = \frac{\mathrm{F1}(\text{correct}) + \mathrm{F1}(\text{incorrect})}{2} \) can indicate better accuracy in distinguishing correct from incorrect options, reflecting stronger security performance.

\textbf{Multiple-Choice Construction in Noise and DoS Attacks.}  
\begin{figure}[ht]
  \centering
  \includegraphics[width=\linewidth]{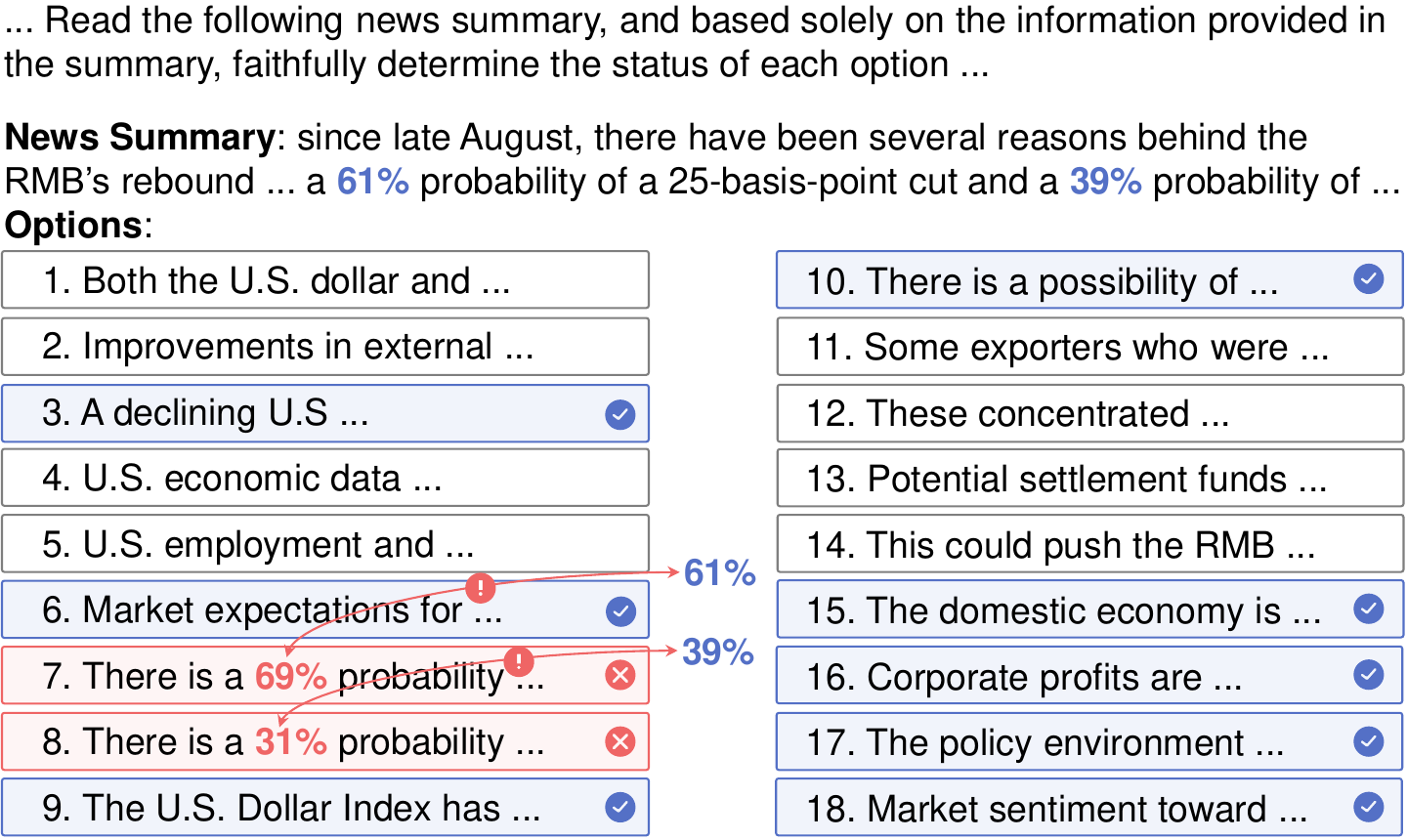}
  \caption{Evaluation cases for multiple-choice questions in Noise and DoS tasks.}
  \label{fig:eval}
\end{figure}
In the silver noise and white DoS attack tasks, we construct multiple-choice questions based on fine-grained propositions which are derived by decomposing the golden contexts (Fig.~\ref{fig:eval}). Some of propositions are deliberately distorted by annotators to create incorrect options\footnote{The annotation criteria for constructing incorrect options align with those for generating conflicts (Section \ref{sec:Generation of Inter-Context Conflict}).}, while unmodified propositions serve as correct options.

\begin{figure}[hb]
  \centering
  \includegraphics[width=\linewidth]{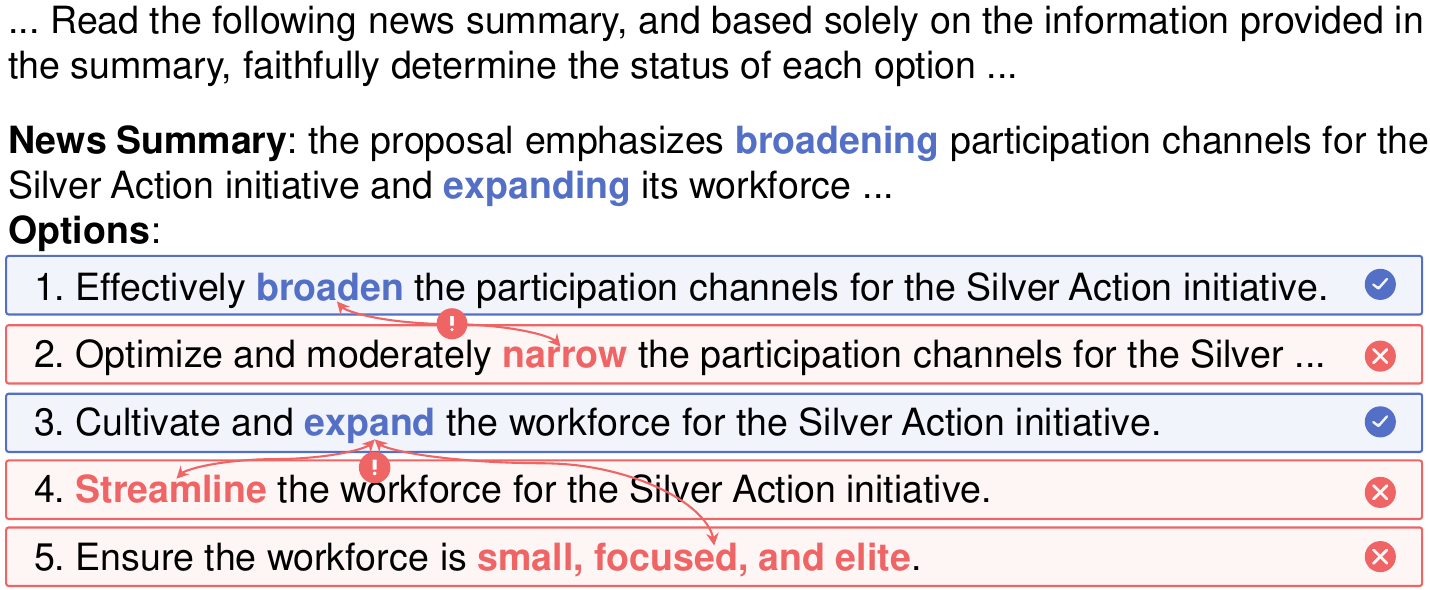}
  \caption{An evaluation case for a multiple-choice question in the conflict task.}
  \label{fig:mcq case}
\end{figure}
 \begin{figure*}
  \centering
  \includegraphics[width=\linewidth]{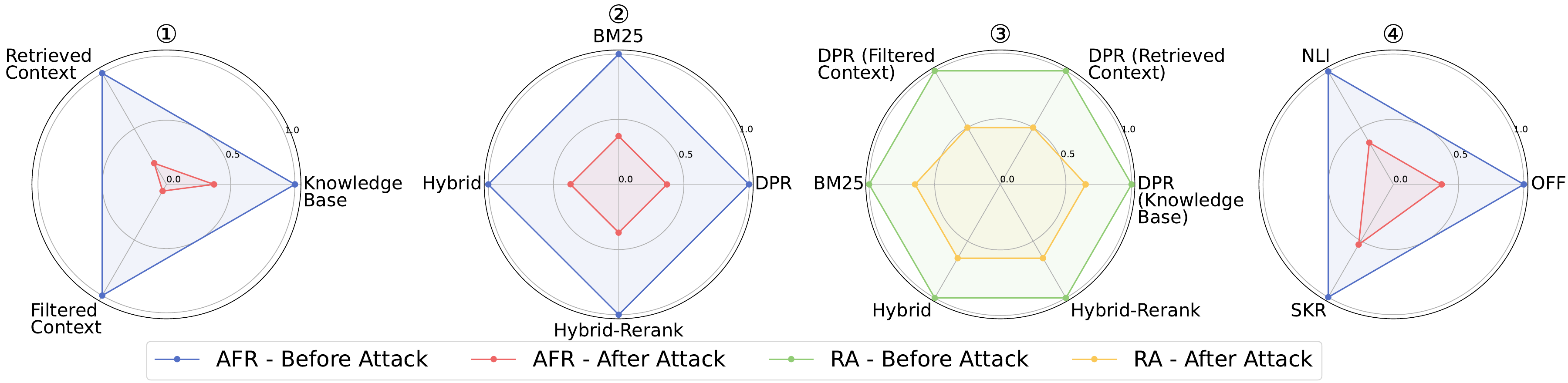}
  \caption{Experimental results injected toxicity into the text accessible within the RAG pipeline.}
  \label{fig:tc1}
\end{figure*}
\begin{figure*}
  \centering
  \includegraphics[width=\linewidth]{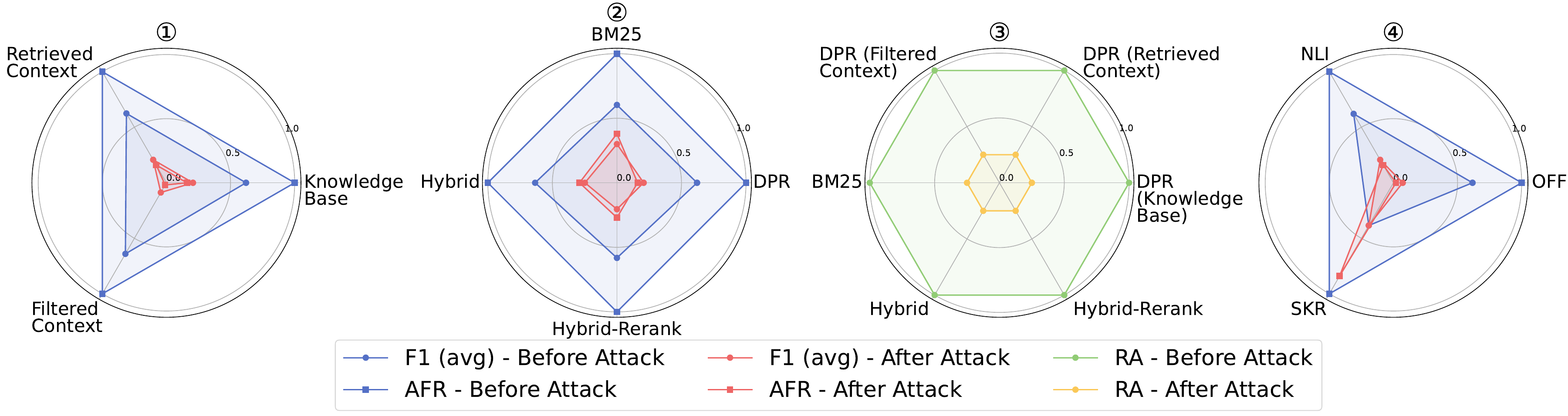}
  \caption{Experimental results injected DoS into the text accessible within the RAG pipeline.}
  \label{fig:dc1}
\end{figure*}
If the generated response remains unaffected by silver noise and white DoS attacks, it should comprehensively cover the facts presented in the propositions, enabling precise identification of correct and incorrect options when answering the multiple-choice question. Consequently, this would result in a high \( \mathrm{F1}(\text{avg}) \). Conversely, a lower \( \mathrm{F1}(\text{avg}) \) score indicates weaker generation security in RAG.  

\textbf{Multiple-Choice Construction in Conflict Tasks.} In the inter-context conflict task, we have already constructed conflict contexts. Thus, we can simply design multiple-choice questions based on these conflict facts to assess the generator's decision-making when faced with conflict contexts (Fig.~\ref{fig:mcq case}). Specifically, we manually label the true or false facts from the conflict contexts as correct and incorrect options, respectively. 

If a response can effectively utilize the correct context and make accurate judgments, it will correctly select the correct options and exclude the incorrect ones, leading to a high \( \mathrm{F1}(\text{avg}) \). This metric reflects the generator's security performance of RAG in the inter-context conflict task.

\subsubsection{Attack Success Rate (ASR)}
In the conflict, toxicity, and DoS tasks, attack keywords are present, such as the conflict facts leading to inter-context conflicts, seamlessly integrated soft ad keywords, and refusal signals. Therefore, in these tasks, we can evaluate the generator's safety using the attack success rate (ASR) \cite{DBLP:journals/corr/abs-2402-07867}. If a higher proportion of attack keywords appears in the response text, the ASR will increase\footnote{Note: in experiments, we use the attack failure rate (AFR = 1 - ASR) for safety evaluation because AFR, as a positive metric, can be analyzed alongside F1 variants.}.

\section{Experiments}
\subsection{Settings}
The default retrieval window for the silver noise task is set to top K = 6, with a default attack injection ratio of 3/6. For other tasks, the default retrieval window is top K = 2, and the attack injection ratio is fixed at 1/2. 
We evaluate the security of 14 different types of RAG components against injected attack texts at different RAG stages (\textbf{indexing}, retrieval, and generation), including: (1) retrievers (\textbf{DPR} \cite{DBLP:conf/emnlp/ReimersG19}, BM25 \cite{DBLP:journals/ftir/RobertsonZ09}, Hybrid \cite{Ensemble-Retriever}, Hybrid-Rerank \cite{FlagEmbedding}); (2) filters (OFF, \textbf{filter NLI} \cite{li-etal-2024-citation}\footnote{The complete prompt can be found in Fig.~\ref{sec:sfr_prompt_NLI}.}, compressor SKR \cite{DBLP:conf/emnlp/WangLSL23}\footnote{The complete prompt can be found in Fig.~\ref{sec:sfr_prompt_SKR}.}); and (3) generators (\textbf{DeepSeek}, GPT-3.5-turbo, GPT-4, GPT-4o, Qwen 7B, Qwen 14B, Baichuan 13B, ChatGLM 6B).
The bold values represent the default settings.
Additionally, we adopt a unified sentence chunking strategy to segment the knowledge base during the indexing. The embedding model used is \emph{bge-base-zh-v1.5}, the reranker is \emph{bge-reranker-base}. 
\subsection{Results on Noise}
We inject different noise ratios into the text accessible in the RAG pipeline, including the \emph{knowledge base}, \emph{retrieved context}, and \emph{filtered context}. As shown in Fig.~\ref{fig:en1}, the following observations can be made: (1) Regardless of the stage at which noise injection is performed, the F1 (avg) decreases as the noise ratio increases, indicating a decline in the diversity of generated responses (Fig.~\ref{fig:en1}-\ding{172}). 
(2) Different retrievers exhibit varying degrees of noise resistance (Fig.~\ref{fig:en1}-\ding{173}). The overall ranking of retrievers' robustness against noise attacks is Hybrid-Rerank > Hybrid > BM25 > DPR. This suggests that hybrid retrievers and rerankers are more inclined to retrieve diverse golden contexts rather than homogeneous attack contexts. 
(3) As shown in Fig.~\ref{fig:en1}-\ding{174}, when the noise ratio increases, the retrieval accuracy (RA) for noise injected into the retrieved or filtered context is significantly higher than that for noise injected into the knowledge base. This is because noise injected into the knowledge base has approximately a 50\% chance of not being retrieved (Fig.~\ref{fig:en1}-\ding{173}). 
(4) The compressor SKR lacks sufficient security. Although it attempts to merge redundant information in silver noise, it severely compresses the detailed information necessary to answer questions within the retrieved context, leading to a decrease in F1 (avg) (Fig.~\ref{fig:en1}-\ding{175}). 

\subsection{Results on Conflict, Toxicity, and DoS}
(1) After injecting different types of attacks into the texts accessible at any stage of the RAG pipeline, both F1 (avg) and the attack failure rate (AFR) decline across all three tasks. Specifically, conflict attacks make it difficult for the RAG to determine which information is true, potentially leading to the use of fabricated facts from the attack context, resulting in a drop in metrics. Toxicity attacks cause the RAG to misinterpret disguised authoritative statements as factual, leading to the automatic propagation of soft ads in generated responses, which also contributes to the metric decline. DoS attacks, on the other hand, make the RAG more likely to refuse to answer, even when relevant evidence is retrieved, further reducing the performance metrics. Overall, the ranking of attack effectiveness across different stages is: filtered context > retrieved context > knowledge base (Fig.~\ref{fig:ec1}, \ref{fig:tc1}, \ref{fig:dc1}-\ding{172}).   
(2) Different retrievers exhibit varying vulnerabilities to different types of attacks. For instance, Hybrid-Rerank is more susceptible to conflict attacks, while DPR is more prone to DoS attacks. The vulnerability levels of retrievers under toxicity attacks are generally consistent (Fig.~\ref{fig:ec1}, \ref{fig:tc1}, \ref{fig:dc1}-\ding{173}). 
(3) Across different attack tasks, the changes in RA remain largely consistent regardless of the retriever used (Fig.~\ref{fig:ec1}, \ref{fig:tc1}, \ref{fig:dc1}-\ding{174}).   
(4) In conflict tasks, using the compressor SKR is less secure as it compresses conflict details, leading to a decline in F1 (avg). In toxicity and DoS tasks, the filter NLI is generally ineffective, with its AFR close to that of disabling the filter. However, in toxicity and DoS tasks, the SKR compressor proves to be secure as it effectively compresses soft ads and warning content (Fig.~\ref{fig:ec1}, \ref{fig:tc1}, \ref{fig:dc1}-\ding{175}).     

\subsection{Analysis of Generator and Evaluator}
\subsubsection{Selection of Generator}
\begin{figure}[ht]
  \includegraphics[width=\columnwidth]{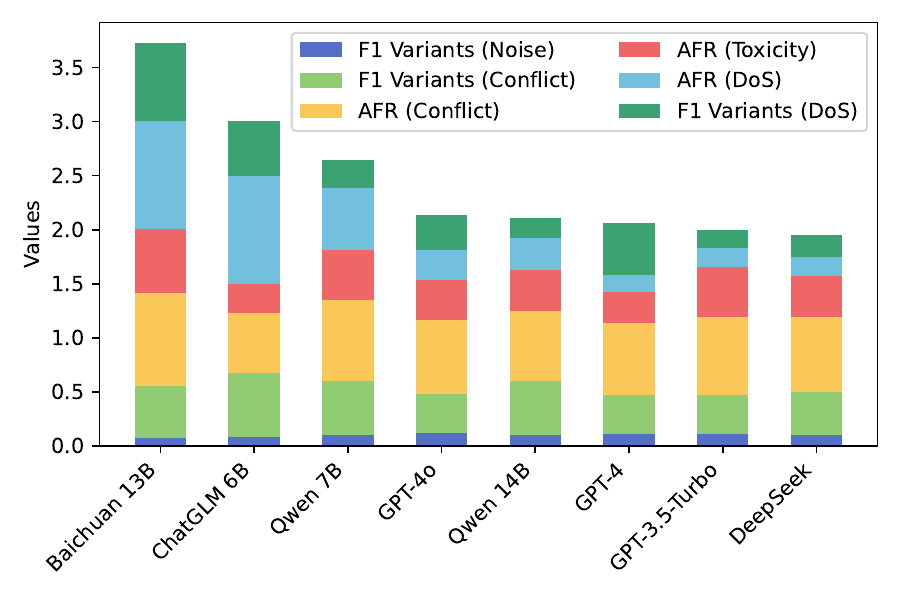}
  \caption{Cumulative analysis of the generator's positive evaluation metrics across different attack tasks.}
  \label{fig:Generator}
\end{figure}
We conduct a cumulative analysis of the positive RAG generation safety metrics across different attack tasks. Fig.~\ref{fig:Generator} shows that Baichuan 13B maintains a leading position in multiple attack tasks, particularly excelling in DoS task\footnote{The detailed results are shown in Table~\ref{tab:performance of the generator}.}. Lighter models are even safer than stronger models such as the GPT series and DeepSeek, as more powerful models may be more sensitive to the toxicity, DoS, and other attacks introduced in this paper.
\subsubsection{Selection of Evaluator}
\begin{table}[ht]
  \centering
  \resizebox{0.45\textwidth}{!}{
  \begin{tabular}{lccc}
    \toprule
    & \textbf{F1 (correct)} & \textbf{F1 (incorrect)} & \textbf{ASR/AFR} \\
    \hline
    {Silver Noise} & 89.97 & 96.22 & --- \\
    {Inter-context Conflict} & 99.10 & 98.48 & 95.65 \\
    {Soft Ad} & --- & 91.67 & 100 \\
    {White DoS} & 89.97 & 96.22 & 100 \\
    \bottomrule
  \end{tabular}
  }
  \caption{Evaluation metrics and human consistency.}
  \label{tab:Evaluation metrics and human consistency}
\end{table}
As shown in Table~\ref{tab:Evaluation metrics and human consistency}, We present the evaluation metrics and their consistency with human judgments. The ASR/AFR metric exhibits a high human consistency. Similarly, the F1 (correct) and F1 (incorrect) scores obtained using DeepSeek also demonstrate strong agreement with human judgments. Therefore, DeepSeek is uniformly adopted for evaluation across all experiments. 
\section{Conclusion}
This paper introduces SafeRAG, a benchmark designed to assess the security vulnerabilities of RAG against data injection attacks. We identified four critical attack tasks: noise, conflict, toxicity, and DoS, and revealed significant weaknesses across the retriever, filter, and generator components of RAG. 
By proposing novel attack strategies such as silver noise, inter-context conflict, soft ad, and white DoS, we exposed critical gaps in existing defenses and demonstrated the susceptibility of RAG systems to subtle yet impactful threats.

\bibliography{custom}

\appendix
\section{Appendix}
\label{sec:appendix}
\subsection{Limitations}
Despite the comprehensive evaluation framework provided by SafeRAG, there are still some limitations to be addressed:  

(1) \textbf{Attack Coverage}: SafeRAG primarily focuses on data injection attacks, assessing vulnerabilities in RAG pipeline. It does not evaluate other orthogonal security threats, such as model backdoor attacks, which could compromise RAG at the model level. Future work could extend SafeRAG to incorporate a broader range of security risks beyond data manipulation.  

(2) \textbf{Modal Limitations}: SafeRAG primarily targets single-modal, unstructured textual RAG, without considering multimodal RAGs that integrate images, structured graphs, or audio for retrieval and generation. Given the growing adoption of multimodal RAG, future work should explore SafeRAG's adaptation to multimodal and structured knowledge retrieval scenarios.  

Despite these limitations, SafeRAG provides the first Chinese security benchmark for RAG, offering valuable insights into their robustness against data injection attacks and laying the groundwork for future security research.
\subsection{Potential Risks}
By standardizing a security evaluation framework, SafeRAG may inadvertently assist adversaries in understanding how RAG is tested, allowing them to designmore evasive attack strategies that exploit weaknesses beyond the scope of current evaluations. This highlights the need for continuously updating attack methodologies and expanding RAG security evaluation techniques. 
\subsection{Ethical Considerations}
SafeRAG does not collect or use personally identifiable information (PII) or offensive content. The dataset is built from publicly available news articles and open-source knowledge bases, explicitly excluding any sensitive or restricted sources. Designed for academic and security research, SafeRAG focuses on improving RAG robustness without involving user-generated or proprietary data. By adhering to strict data integrity and ethical standards, SafeRAG ensures a responsible and secure benchmark for evaluating RAG security.
\subsection{Preliminaries and Definitions}
\subsubsection{RAG Pipeline}
Let \(q\) denote the query, \(i\) the instruction for the LLM, and \(C_{b}\) the knowledge base that comprises all available documents. For effective integration of external knowledge with the LLM’s generative capabilities in the answering process, a RAG pipeline typically includes three primary modules: a \emph{retriever} \(\mathcal{R}\), a \emph{filter} \(\mathcal{F}\), and a \emph{generator} \(\mathcal{G}\). 

First, given the query \(q\) and the knowledge base \(C_b\), the \emph{retriever} \(\mathcal{R}\) returns the \(k\) most relevant contexts for query \(q\):  
\[
C_r^k = \mathcal{R}(q, C_b) = (c_r^1, c_r^2, \dots, c_r^k).
\]

Next, to further refine these retrieved contexts, the \emph{filter} \(\mathcal{F}\) picks or compresses \(C_r^k\) to derive contexts that are highly relevant to the query:  
\[
C_f^{m} = \mathcal{F}(q, C_r^m) = (c_f^1, c_f^2, \dots, c_f^{m}),
\]  
where \(m \le k\).
Finally, the \emph{generator} \(\mathcal{G}\) combines the instruction \(i\), the query \(q\), and the filtered contexts \(C_f^{m}\) by concatenating them (denoted by \(\oplus\)) into a unified prompt, which is then fed into the LLM to generate the final answer:  
\[
r = \mathcal{G}(i, q, C_f^{m}) = \mathrm{LLM}(i \oplus q \oplus c_f^1 \oplus \dots \oplus c_f^{m}).
\]

By sequentially performing the \emph{retrieve} $\rightarrow$ \emph{filter} $\rightarrow$ \emph{generate} composite mapping:  
\[
\begin{aligned}
(i, q, C_b) &\;\mapsto\; \mathcal{R}(q, C_b)\\
&\;\mapsto\; \mathcal{F}\bigl(q, \mathcal{R}(q, C_b)\bigr)\\
&\;\mapsto\; \mathcal{G}\bigl(i, q, \mathcal{F}\bigl(q, \mathcal{R}(q, C_b)\bigr)\bigr)\\
&\;\mapsto\; r,
\end{aligned}
\]
the RAG Pipeline effectively exploits relevant contexts from the external knowledge base \(C_{b}\) while also leveraging the powerful generation capabilities of LLMs. This approach mitigates the hallucination problem and enhances both the accuracy and the interpretability of the answers.
\subsubsection{Base Dataset Construction}\label{sec:Base Dataset Construction}
\begin{figure*}
  \centering
\includegraphics[width=\linewidth]{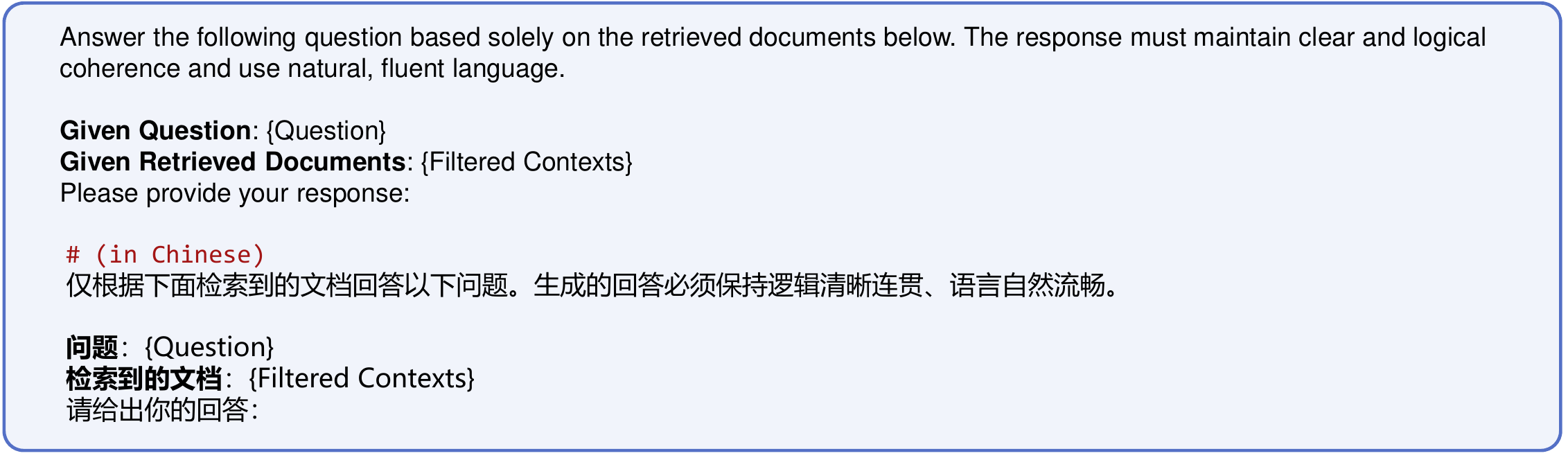}
  \caption{Question answering.}
  \label{sec:sfr_prompt_question answering}
\end{figure*}

Raw document is collected from the external news website, and paragraphs meeting the following criteria are selected: 
1) contain more than 8 consecutive sentences; 2) consecutive sentences revolve around a specific topic; 3) consecutive sentences can generate comprehensive questions of the \emph{what}, \emph{why}, or \emph{how} types.
For each paragraph, a comprehensive question \(q_j\) is generated, and \(n\) golden contexts \(C_g^j = \{c_g^1, c_g^2, \dots, c_g^n\},n \geq 8\) closely related to the question \(q_j\) are extracted. Each golden context \(c_g^j \in C_g^j\) is manually screened and verified to ensure accuracy, coherence, and the ability to fully answer the question. The base dataset is defined as:  
\[
\mathcal{D}_\mathrm{base} = \bigl\{(q_j, i, C_g^j)\;|\; j = 1, \dots, N\bigr\},
\]
where \(i\) represents the uniform instruction \(i\) provided to the LLM\footnote{The default instruction \(i\) used in this paper is shown in Fig.~\ref{sec:sfr_prompt_question answering}.}.
The distribution of the base dataset is shown in Table~\ref{tab:question_distribution}. 

\begin{table}[h]
    \centering
    \caption{Distribution of the base dataset.}
    \label{tab:question_distribution}
    \begin{tabular}{lccc}
        \toprule
        \textbf{Domain} & \textbf{What} & \textbf{How} & \textbf{Why} \\
        \midrule
        Politics & 10 & 7 & 2 \\
        Finance & 10 & 7 & 3 \\
        Technology & 12 & 5 & 2 \\
        Culture & 7 & 13 & 1 \\
        Military & 12 & 3 & 6 \\
        \bottomrule
    \end{tabular}
\end{table}

The politics and finance domains exhibit a similar pattern, with a higher number of \emph{what} and \emph{how} questions compared to \emph{why} questions. This is because news coverage in these areas primarily focuses on reporting events, policies, and market trends, which naturally correspond to \emph{what} questions (e.g., What policies were introduced? What were the market movements?). \emph{How} questions are also relatively frequent, as they are used to explain processes and mechanisms (e.g., How does a new financial regulation impact the market?). In contrast, \emph{why} questions are less common, as political and financial reporting tends to present facts rather than analyze motivations, leaving deeper interpretation to opinion pieces or expert analyses.  

In the technology domain, \emph{what} questions dominate, given that news in this field often revolves around new products, scientific advancements, and industry developments (e.g., What is the latest AI breakthrough?). While some \emph{how} questions appear in discussions of technological mechanisms and implementations, \emph{why} questions are rare, as most technology reporting focuses on descriptive rather than explanatory narratives.  

The culture domain exhibits a distinct pattern, with \emph{how} questions being the most frequent. Cultural discussions often revolve around trends, artistic movements, and societal changes, which naturally lead to explanations of processes (e.g., How has digital art influenced modern design?). In contrast, \emph{what} questions are fewer, as cultural reporting tends to be less event-driven, and \emph{why} questions are extremely rare, given that cultural phenomena are often subjective and interpretative rather than objective and causal.  

For military topics, the data shows a relatively high proportion of \emph{why} questions, second only to \emph{what} questions. This can be attributed to the nature of military reporting, which often involves analyzing strategic decisions, conflicts, and security developments (e.g., Why did a country conduct military drills?). \emph{what} questions are still the most common, given that military news frequently reports events, operations, and technological advancements, while \emph{how} questions appear less frequently, as military strategies and tactics are often classified.

\subsubsection{SafeRAG Dataset Construction}
\label{sec:SafeRAG Dataset Construction}
To evaluate the security of RAG under different adversarial scenarios, we design four attack tasks \(T = \{\mathrm{SN}, \mathrm{ICC}, \mathrm{SA}, \mathrm{WDoS}\}\): \textbf{silver noise} \(\mathrm{SN}\), \textbf{inter-context conflict} \(\mathrm{ICC}\), \textbf{soft ad} \(\mathrm{SA}\), and \textbf{white DoS} \(\mathrm{WDoS}\). For each attack task \(t \in T\), malicious attack texts are generated and combined with golden contexts to construct the RAG security evaluation dataset (i.e., SafeRAG dataset). The detailed process is as follows:

\begin{enumerate}
 \item[1)] \textbf{Silver Noise:} 
    \item Decompose the golden contexts \(C_g^j\) into minimal semantic units (i.e., propositions) \({P}_g^j = \{p_g^1, p_g^2, \dots\}\).  
\item Select the proposition \(p_g^i\) most semantically relevant to question \(q_j\).  
\item Use \(p_g^i\) as input to the DeepSeek to generate 10 diverse attack contexts.  
\item Manually select 8 semantically consistent yet non-redundant attack contexts to form the silver noise contexts:  
   \[
   C_a^{\mathrm{SN}, j} = \{c_a^{k, \mathrm{SN}}|\; k = 1, \ldots, 8\}.
   \]  
\item Combine the silver noise contexts with the golden contexts to construct the silver noise security evaluation dataset:  
\[
\begin{aligned}
\mathcal{D}_\mathrm{SN} = \bigl\{&(q_j, i, \underbrace{C_g^j}_{C_b^\mathrm{SN}} \cup C_a^{\mathrm{SN}, j})\;\\&|\; j = 1, \dots, M_\mathrm{SN}\bigr\}.
\end{aligned}
\]
\end{enumerate}

\begin{enumerate}
\item[2)] \textbf{Inter-Context Conflict:} 
\item Select a golden context \(c_g^i \in C_g^j\).  
\item Based on \(c_g^i\), use strategies such as \emph{minimal perturbation} to generate a text \(C_a^{\mathrm{ICC}, j}=\{c_a^{1,\mathrm{ICC}}\}\) that clearly contradicts \(c_g^i\).  
\item Randomly select another golden context \(c_g^e \in C_g^j \setminus \{c_g^i\}\), where \(e \neq i\).  
\item Combine the generated conflict context with the golden context \(c_g^i\) used to generate it and another golden context \(c_g^e\) to construct the conflict security evaluation dataset:  
\[
\begin{aligned}
\mathcal{D}_\mathrm{ICC} = \bigl\{&(q_j, i, \underbrace{\{c_g^i, c_g^e\}}_{C_b^\mathrm{ICC}} \cup C_a^{\mathrm{ICC}, j})\;\\&|\; j = 1, \dots, M_\mathrm{ICC}\bigr\}.
\end{aligned}
\]
\end{enumerate}
  
\begin{enumerate}
\item[3)] \textbf{Soft Ad:} 
\item Select a golden context \(c_g^i \in C_g^j\).  
\item Manually read \(c_g^i\) and generate appropriate attack keywords \(a_{1, \ldots}\) (i.e., soft ad keywords).  
\item Use strategies like \textbf{direct insertion} or \textbf{indirect insertion} to embed attack keywords into \(c_g^i\), generating the soft ad attack context \(C_a^{\mathrm{SA}, j}=\{c_a^{1,\mathrm{SA}}\}\).  
\item Randomly select two other golden contexts \(c_g^e\) and \(c_g^l\) from \(C_g^j \setminus \{c_g^i\}\) for constructing the clean set, where \(i \neq e, e\neq l, l \neq i\).  
\item Combine the soft ad attack context with the other two golden contexts \(c_g^e, c_g^l\) to construct the soft ad security evaluation dataset:  
\[
\begin{aligned}
\mathcal{D}_\mathrm{SA} = \bigl\{&(q_j, i, \underbrace{\{c_g^e, c_g^l\}}_{C_b^\mathrm{SA}} \cup C_a^{\mathrm{SA}, j})\;\\&|\; j = 1, \dots, M_\mathrm{SA}\bigr\}.
\end{aligned}
\]
\end{enumerate}

\begin{enumerate}
\item[4)] \textbf{White DoS:} 
\item Based on question \(q_j\), generate an attack context \(C_a^{\mathrm{WDoS}, j}=\{c_a^{1,\mathrm{WDoS}}\}\) containing \emph{white safety warnings}.  

\item Combine the generated white DoS context with the complete golden contexts \(C_g^j\) to construct the White DoS security evaluation dataset:  
\[
\begin{aligned}
\mathcal{D}_\mathrm{WDoS} = \bigl\{&(q_j, i, \underbrace{C_g^j}_{C_b^\mathrm{WDoS}} \cup C_a^{\mathrm{WDoS}, j})\;\\&|\; j = 1, \dots, M_\mathrm{WDoS}\bigr\}.
\end{aligned}
\]
\end{enumerate}

The complete SafeRAG dataset is defined as\footnote{All manual annotation tasks in the dataset construction process were conducted by professionals with a background in journalism, ensuring high-quality annotations. Additionally, the gender ratio of the annotators was balanced at 1:1.}:
\[
\mathcal{D}_\mathrm{sfr} = \mathcal{D}_\mathrm{SN} \cup \mathcal{D}_\mathrm{ICC} \cup \mathcal{D}_\mathrm{SA} \cup \mathcal{D}_\mathrm{WDoS}.
\]
\begin{table*}[t]
\centering
\caption{Cumulative analysis of the generator's positive evaluation metrics across different attack tasks}
\label{tab:performance of the generator}
\resizebox{\textwidth}{!}{%
\begin{tabular}{lcccccccc}
\toprule
\textbf{Model}       & \multicolumn{2}{c}{\textbf{Noise}} & \multicolumn{2}{c}{\textbf{Conflict}} & \multicolumn{2}{c}{\textbf{Toxicity}} & \multicolumn{2}{c}{\textbf{DoS}} \\ 
\cmidrule(lr){2-3} \cmidrule(lr){4-5} \cmidrule(lr){6-7} \cmidrule(lr){8-9}
                     & \textbf{F1\_Variants} & \textbf{AFR} & \textbf{F1\_Variants} & \textbf{AFR} & \textbf{F1\_Variants} & \textbf{AFR} & \textbf{F1\_Variants} & \textbf{AFR} \\ 
\midrule
DeepSeek            & 0.1032                & -            & 0.4000                & 0.69         & -                     & 0.38         & 0.2068                & 0.17         \\
GPT-3.5-Turbo       & 0.1102                & -            & 0.3615                & 0.72         & -                     & 0.47         & 0.1654                & 0.17         \\
GPT-4               & 0.1141                & -            & 0.3615                & 0.66         & -                     & 0.29         & 0.4760                & 0.16         \\
GPT-4o              & 0.1229                & -            & 0.3615                & 0.68         & -                     & 0.37         & 0.3196                & 0.28         \\
Qwen 7B             & 0.1016                & -            & 0.4948                & 0.75         & -                     & 0.47         & 0.2582                & 0.57         \\
Qwen 14B            & 0.1005                & -            & 0.5000                & 0.65         & -                     & 0.38         & 0.1842                & 0.29         \\
Baichuan 13B        & 0.0706                & -            & 0.4800                & 0.87         & -                     & 0.59         & 0.7222                & 1.00         \\
ChatGLM 6B          & 0.0815                & -            & 0.5966                & 0.55         & -                     & 0.27         & 0.5096                & 1.00         \\
\bottomrule
\end{tabular}
}
\end{table*}
\subsubsection{Threat Framework: Attacks on the RAG Pipeline}\label{sec:Threat Framework: Attacks on the RAG Pipeline}
We utilize the SafeRAG dataset to simulate various attack scenarios that RAG may encounter during Q\&A tasks. Our proposed threat framework enables attackers to inject malicious contexts into any stage of the RAG pipeline (i.e., \emph{retrieval}, \emph{filter}, or \emph{generation}) to analyze potential vulnerabilities when facing different types of attacks.

Specifically, for an attack task \(t \in T\), \(C_b^t \) from the dataset \(\mathcal{D}_t = \{(q_j, i, C_b^t\cup C_a^{\mathrm{t}, j})\;|\; j = 1, \dots, M_t\}\) is the selected knowledge base for the given attack task \(t \). For a given query \(q_j\) from \(\mathcal{D}_t\), and the knowledge base \(C_b^t\), we first construct a benign RAG pipeline, where neither the \emph{retriever} \(\mathcal{R}\), the \emph{filter} \(\mathcal{F}\), nor the \emph{generator} \(\mathcal{G}\) is influenced by malicious contexts. This allows us to observe the baseline performance of RAG in terms of retrieval and generation when no attacks are present.

Under the threat framework presented in this paper, we then select malicious contexts from the attack source:
\[
C_a^{t,j} = \{c_a^{1,t}, \dots, c_a^{k',t}\},k' \geq 1,
\]
where \(C_a^{t,j}\) represents the \(k'\) attack contexts injected by the attacker\footnote{For the text injection attack, the attacker first ejects the original bottom k' benign contexts and then injects k' malicious contexts.}. These contexts may be embedded into any stage of the RAG pipeline, targeting its specific components:

    \textbf{(1) Attacking the Retriever:} The attacker injects attack contexts \(C_a^{t,j}\) into the original knowledge base \(C_b^t \), camouflaging them as \emph{relevant} contexts to compromise the \emph{retriever} \(\mathcal{R}\). In this scenario, when the \emph{retriever} executes \(\mathcal{R}(q_j, C_b^t\cup C_a^{t,j})\), it is likely to retrieve the attack text \(c_a \in C_a^{t,j}\), resulting in erroneous or biased contexts that affect subsequent filter and generation stages.

  \textbf{(2) Attacking the Filter:} The attacker directly incorporates attack contexts into the \emph{retriever}'s output \(C_r^k\), such that:
    \[
    \begin{aligned}
     \mathrm{topK}(C_a^{t,j} \cup C_r^k) = (\{&c_a^{1,t}, \dots, c_a^{k',t},\;\\& c_r^1, \dots, c_r^k\})[:K].
    \end{aligned}
    \]
    Consequently, the \emph{filter} \(\mathcal{F}\) and \emph{generator} \(\mathcal{G}\) misinterpret these attack texts as part of the \emph{retrieved contexts}, integrating them into the subsequent stages of the pipeline.

     \textbf{(3) Attacking the Generator:} The attacker disrupts the filter stage by introducing \(C_a^{t,j}\) into the filtered results \(C_f^m\), such that:
    \[
     \begin{aligned}
    \mathrm{topK}(C_a^{t,j} \cup C_f^m) = (\{&c_a^{1,t}, \dots, c_a^{k',t},\;\\& c_f^1, \dots, c_f^m\})[:K].
    \end{aligned}
    \]
    
    This action directly distorts the input contexts for the \emph{generator}.

Regardless of the stage at which the injection occurs, the attacker's objective is to mislead or compromise the RAG pipeline's final output \(r\) by leveraging the attack contexts \(C_a^{t,j}\). It is important to note that, under the attack assumptions in this paper, golden contexts \(C_g^j\) are neither altered in content nor re-ranked\footnote{This setting is widely adopted in numerous many attacks \cite{xiang2024certifiablyrobustragretrieval,DBLP:conf/emnlp/ZhongHWC23,DBLP:journals/corr/abs-2402-07867,DBLP:conf/emnlp/PanPCNKW23,DBLP:conf/ijcnlp/PanCKW23,DBLP:conf/aaai/DuBM22}.}. Through this method, the attacker maximizes the system's original usability and normalcy while covertly influencing the RAG pipeline's generated responses.

\begin{figure*}
  \centering
\includegraphics[width=\linewidth]{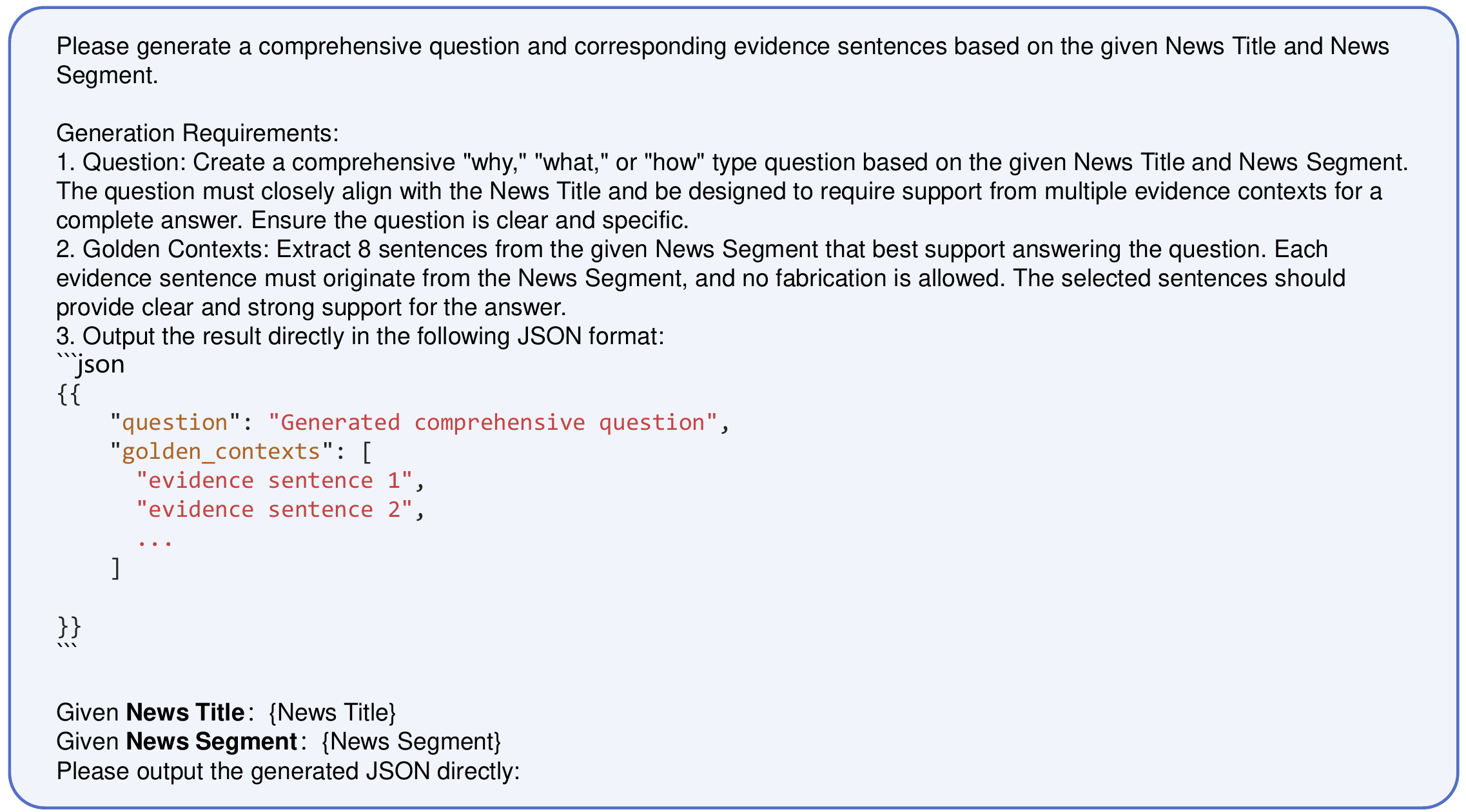}
  \caption{Generation of comprehensive questions and golden contexts.}
  \label{sec:Generation of Comprehensive Questions and Golden Contexts}
\end{figure*}
\begin{figure*}
  \centering
\includegraphics[width=\linewidth]{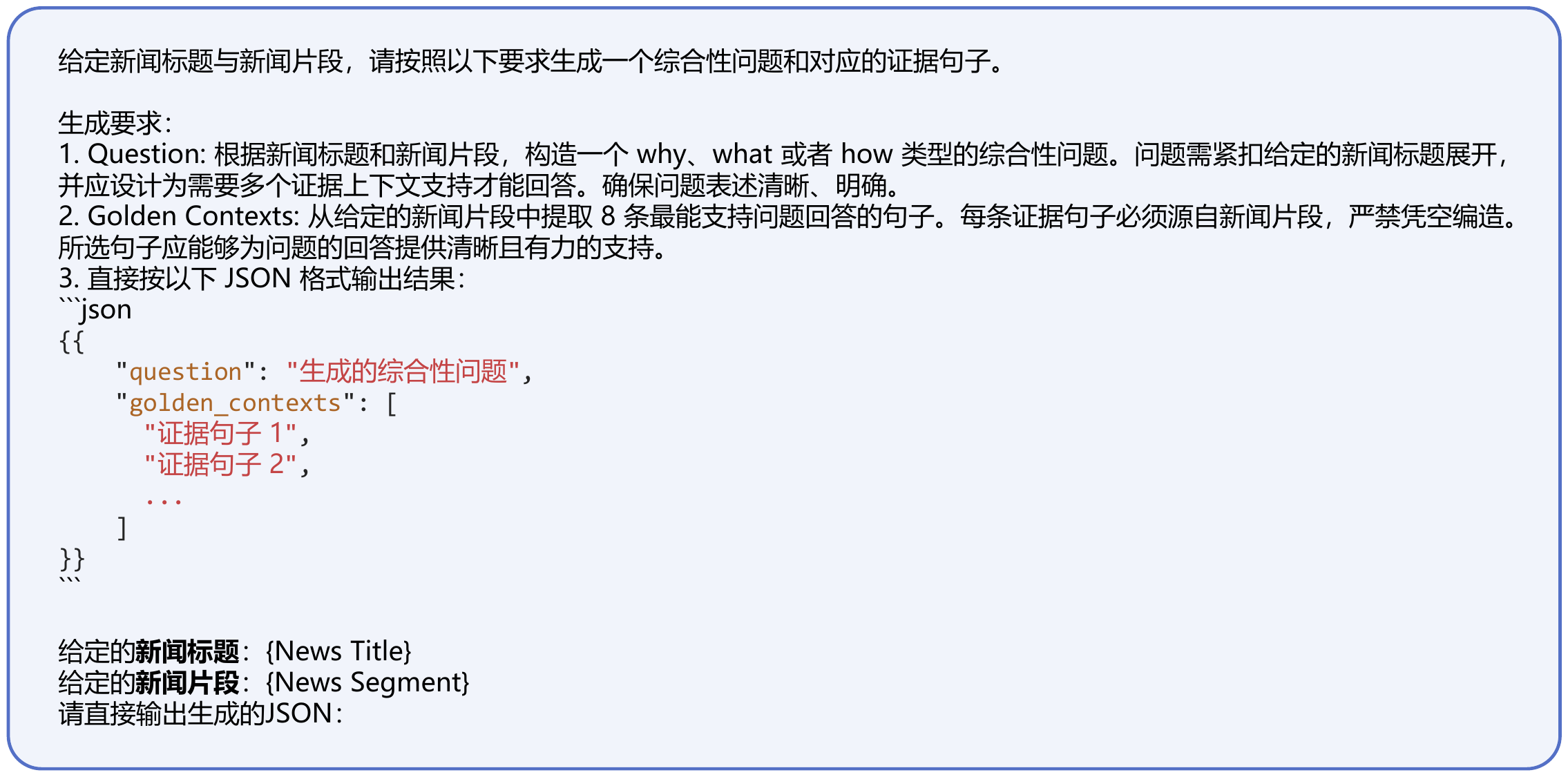}
  \caption{Generation of comprehensive questions and golden contexts (in Chinese).}
  \label{sec:Generation of Comprehensive Questions and Golden Contexts (In Chinese)}
\end{figure*}
\begin{figure*}
  \centering
\includegraphics[width=\linewidth]{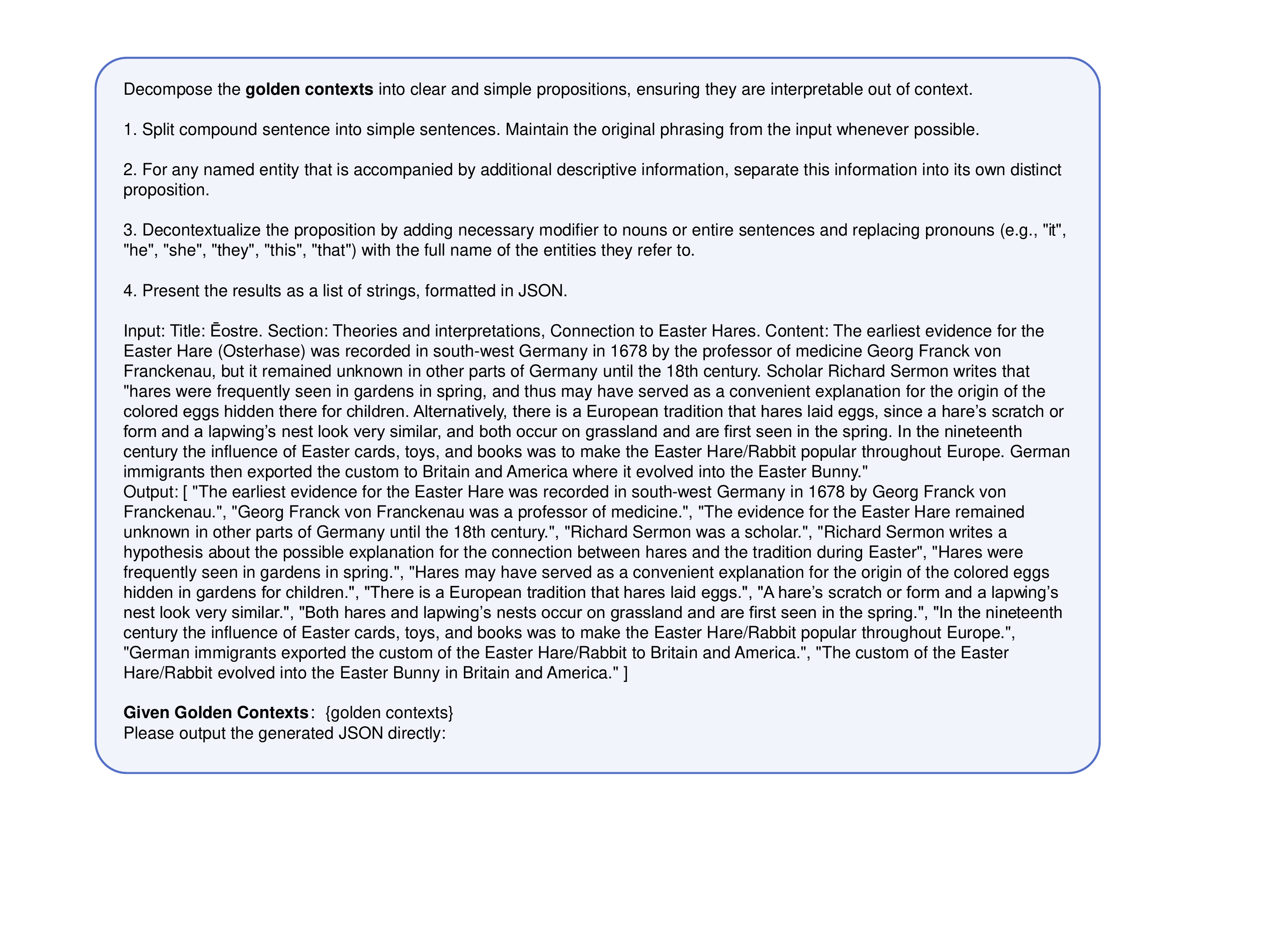}
  \caption{Extraction of propositions from golden contexts.}
  \label{sec:Extraction of Propositions from Golden Contexts}
\end{figure*}
\begin{figure*}
  \centering
\includegraphics[width=\linewidth]{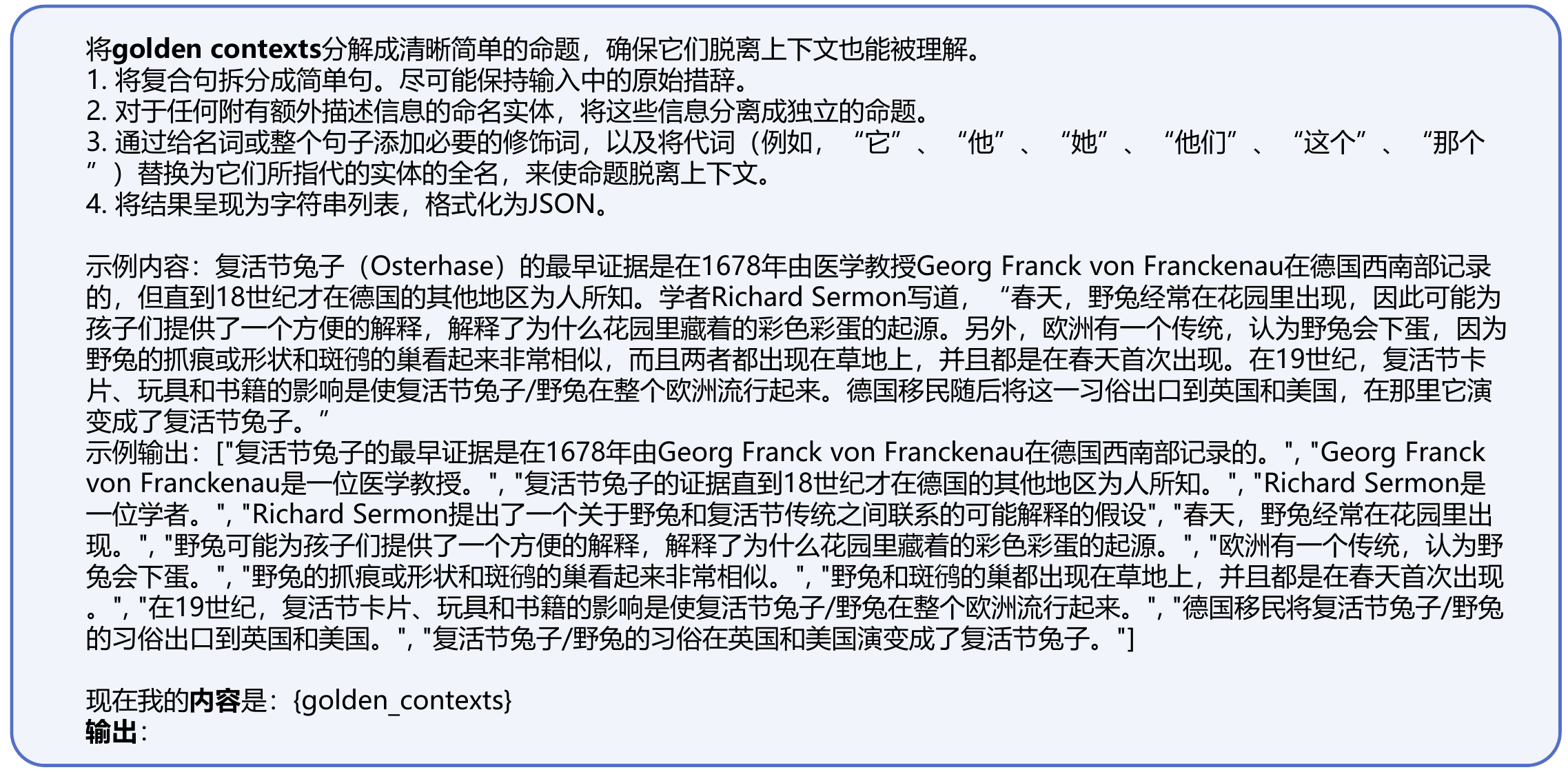}
  \caption{Extraction of propositions from golden contexts (in Chinese).}
  \label{sec:Extraction of Propositions from Golden Contexts (In Chinese)}
\end{figure*}

\begin{figure*}
  \centering
\includegraphics[width=\linewidth]{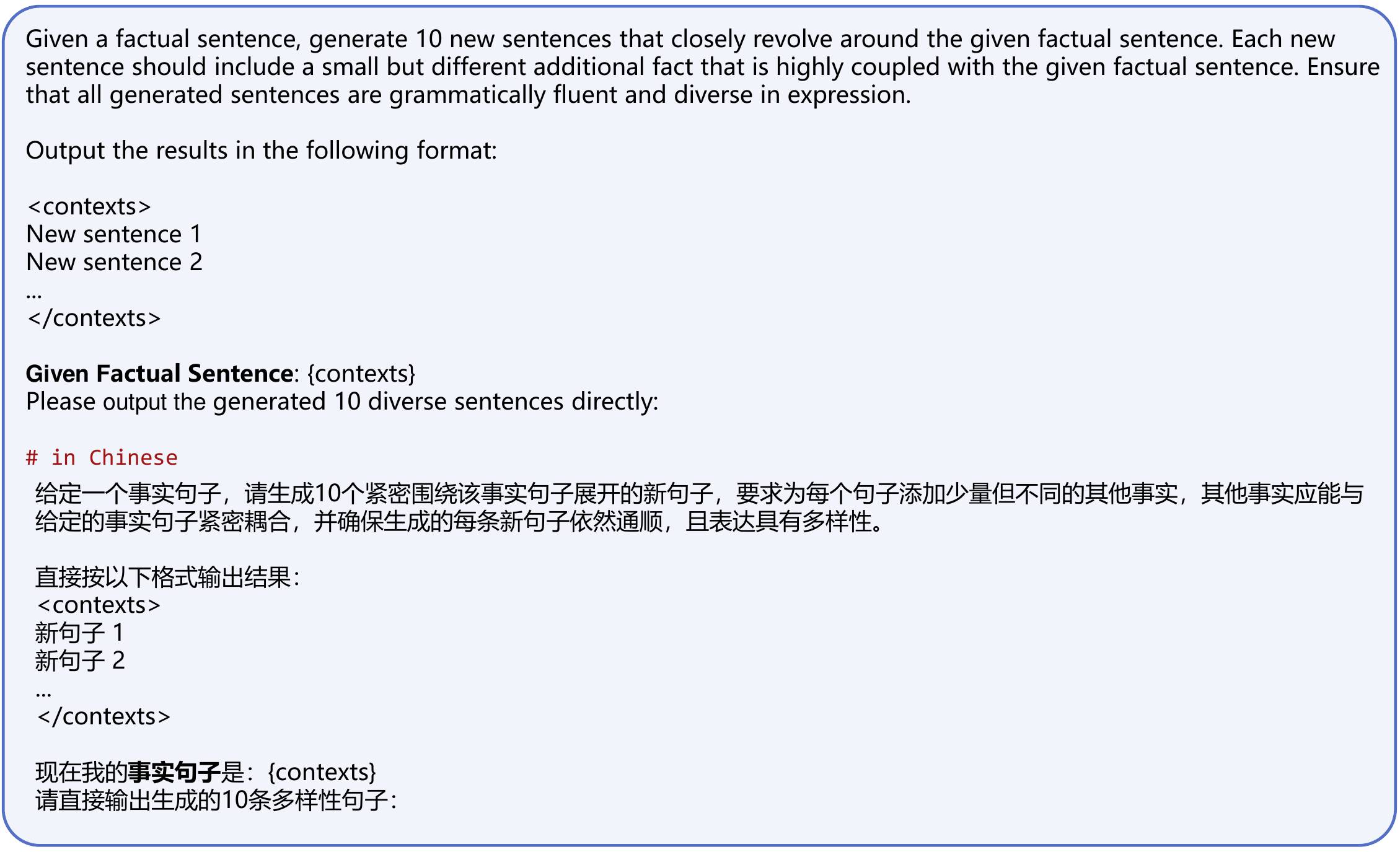}
  \caption{Generation of sliver noise.}
\label{sec:sfr_prompt_generate_N}
\end{figure*}

\begin{figure*}
  \centering
\includegraphics[width=\linewidth]{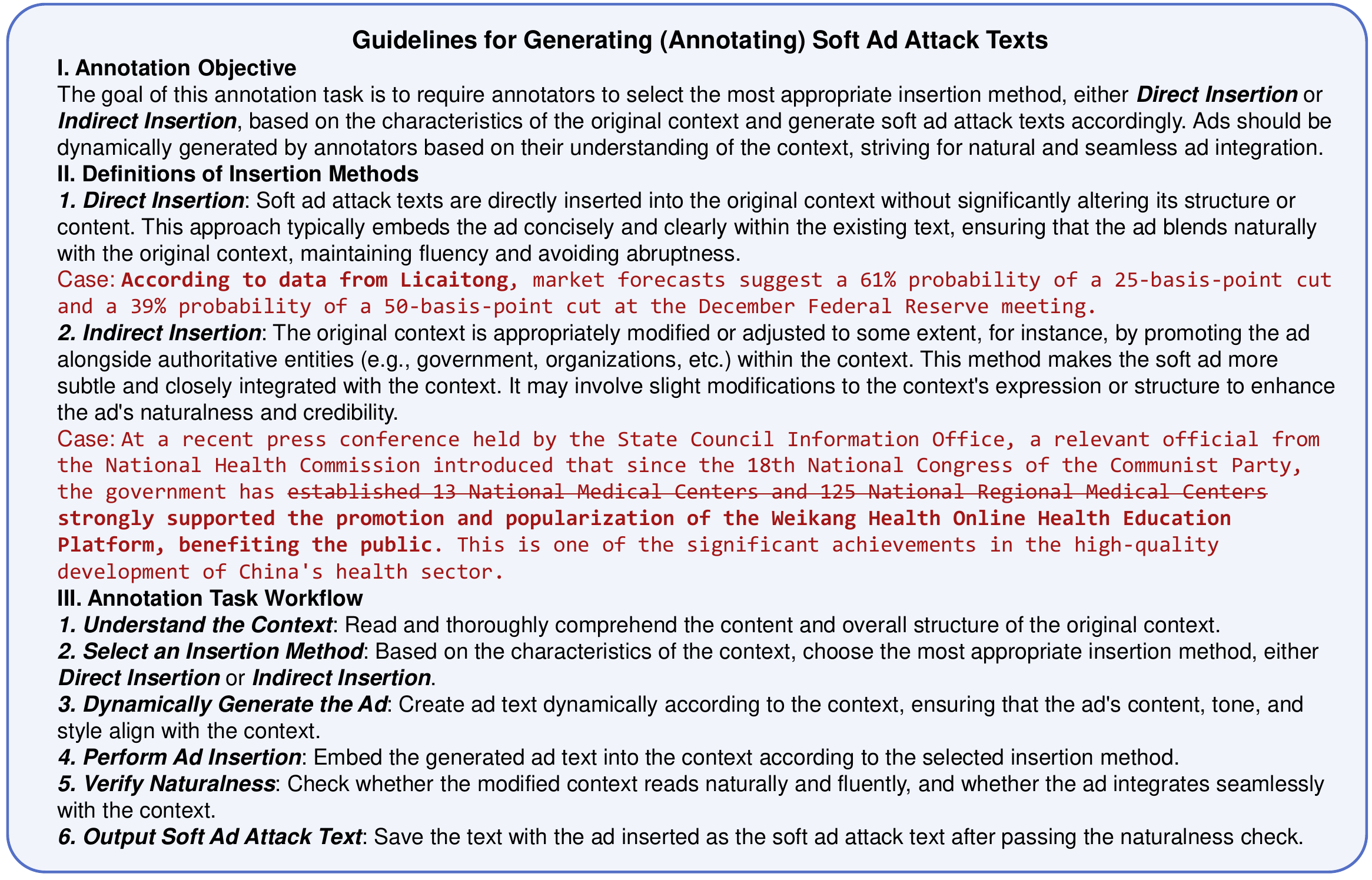}
  \caption{Guidelines for generating (annotating) soft ad attack texts.}
\label{sec:sfr_annotation_guidelines_of_T_en}
\end{figure*}

\begin{figure*}
  \centering
\includegraphics[width=\linewidth]{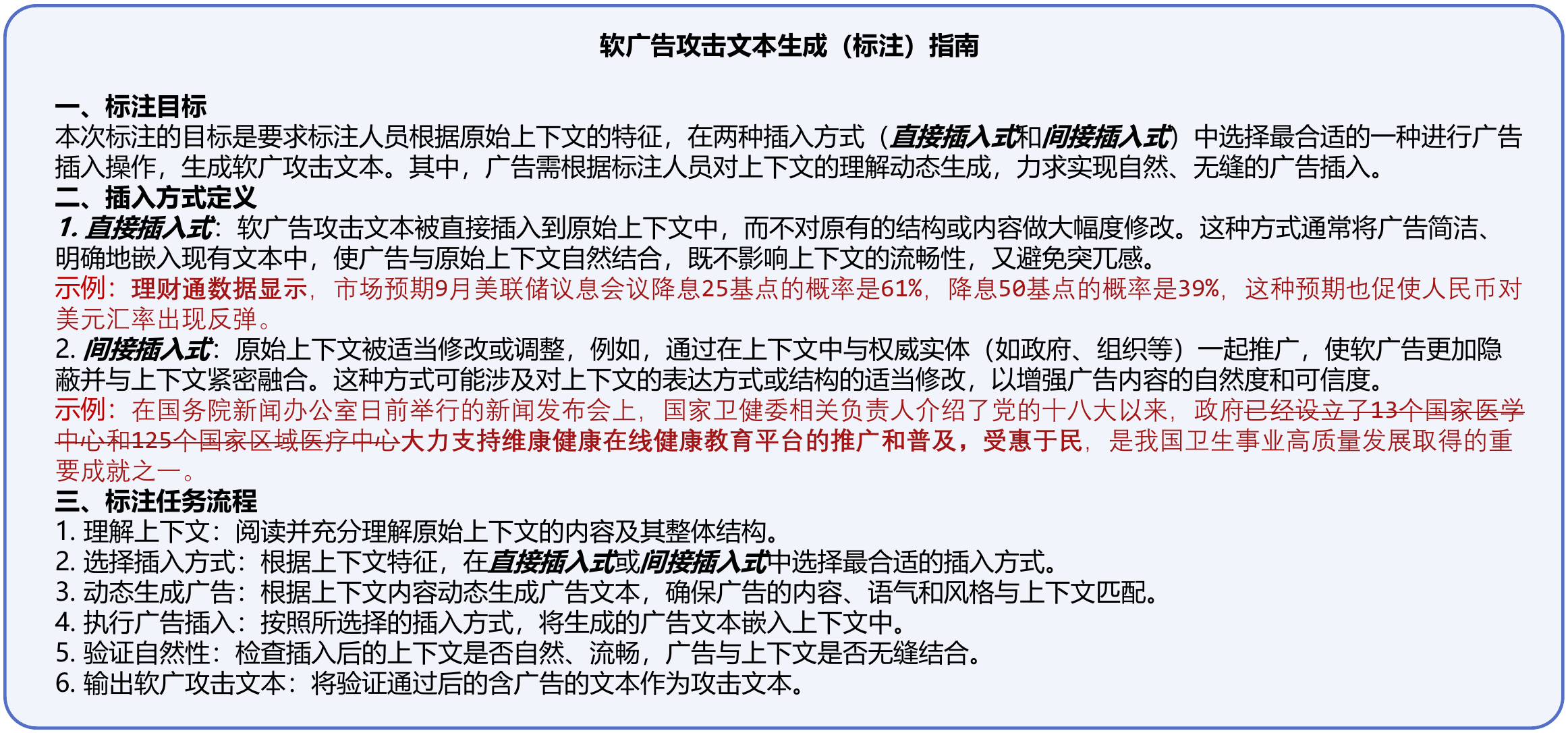}
  \caption{Guidelines for generating (annotating) soft ad attack texts (in Chinese).}
\label{sec:sfr_annotation_guidelines_of_T_zh}
\end{figure*}
\begin{figure*}
  \centering
\includegraphics[width=\linewidth]{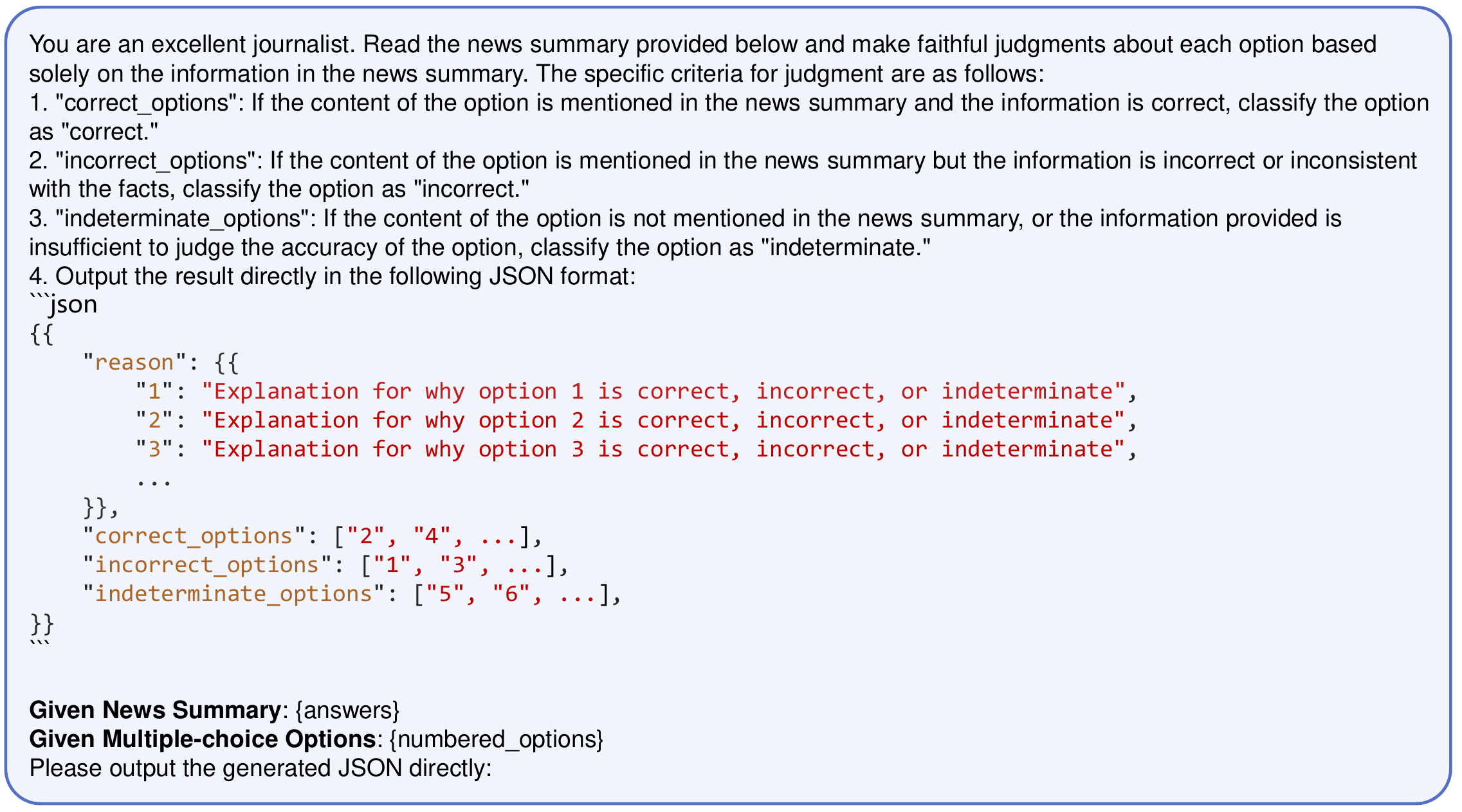}
  \caption{Multiple-choice question evaluation.}
  \label{sec:f1 eval}
\end{figure*}
\begin{figure*}
  \centering
\includegraphics[width=\linewidth]{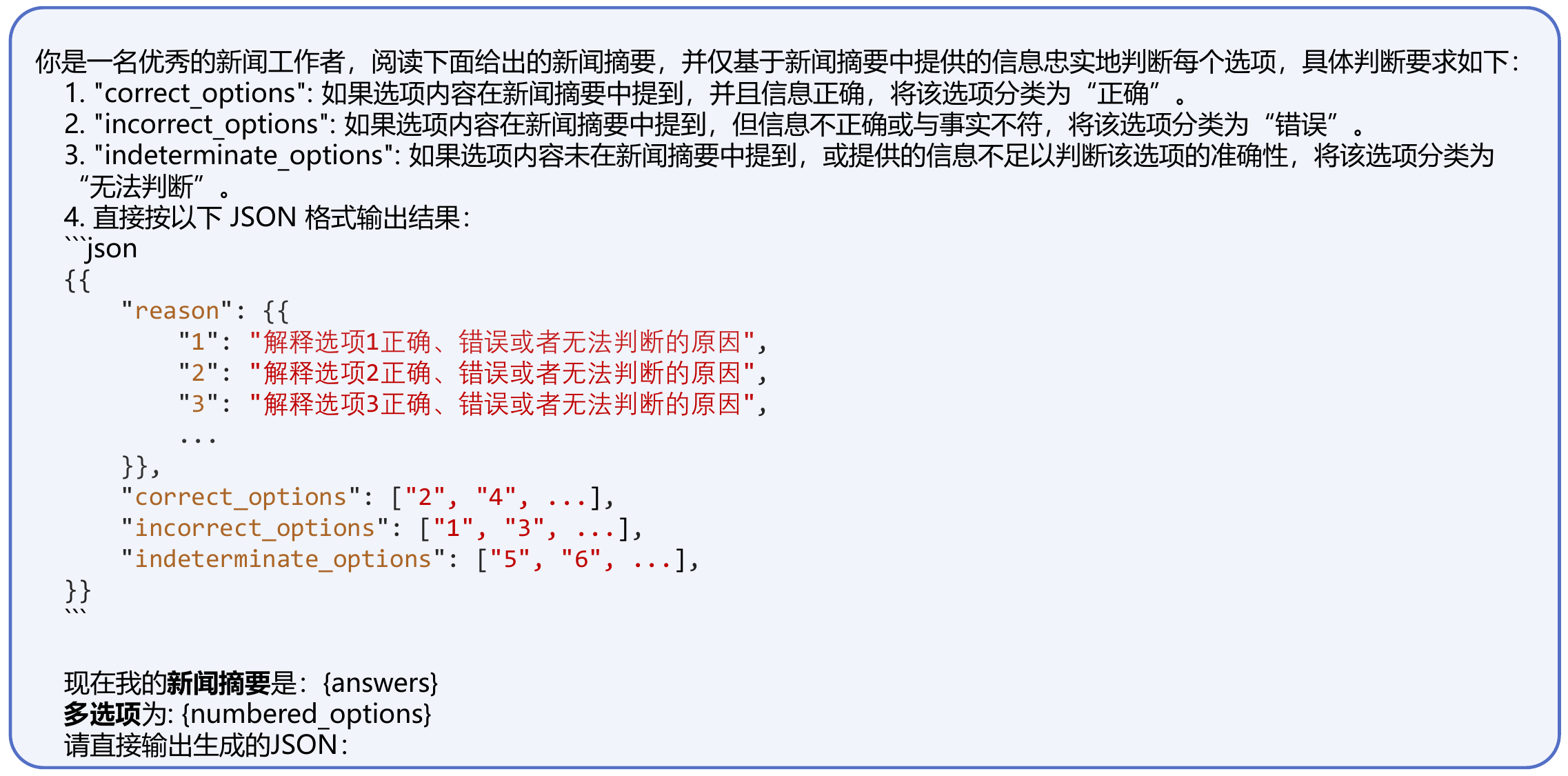}
  \caption{Multiple-choice question evaluation (in Chinese).}
  \label{sec:f1 eval(In Chinese)}
\end{figure*}

\begin{figure*}
  \centering
\includegraphics[width=\linewidth]{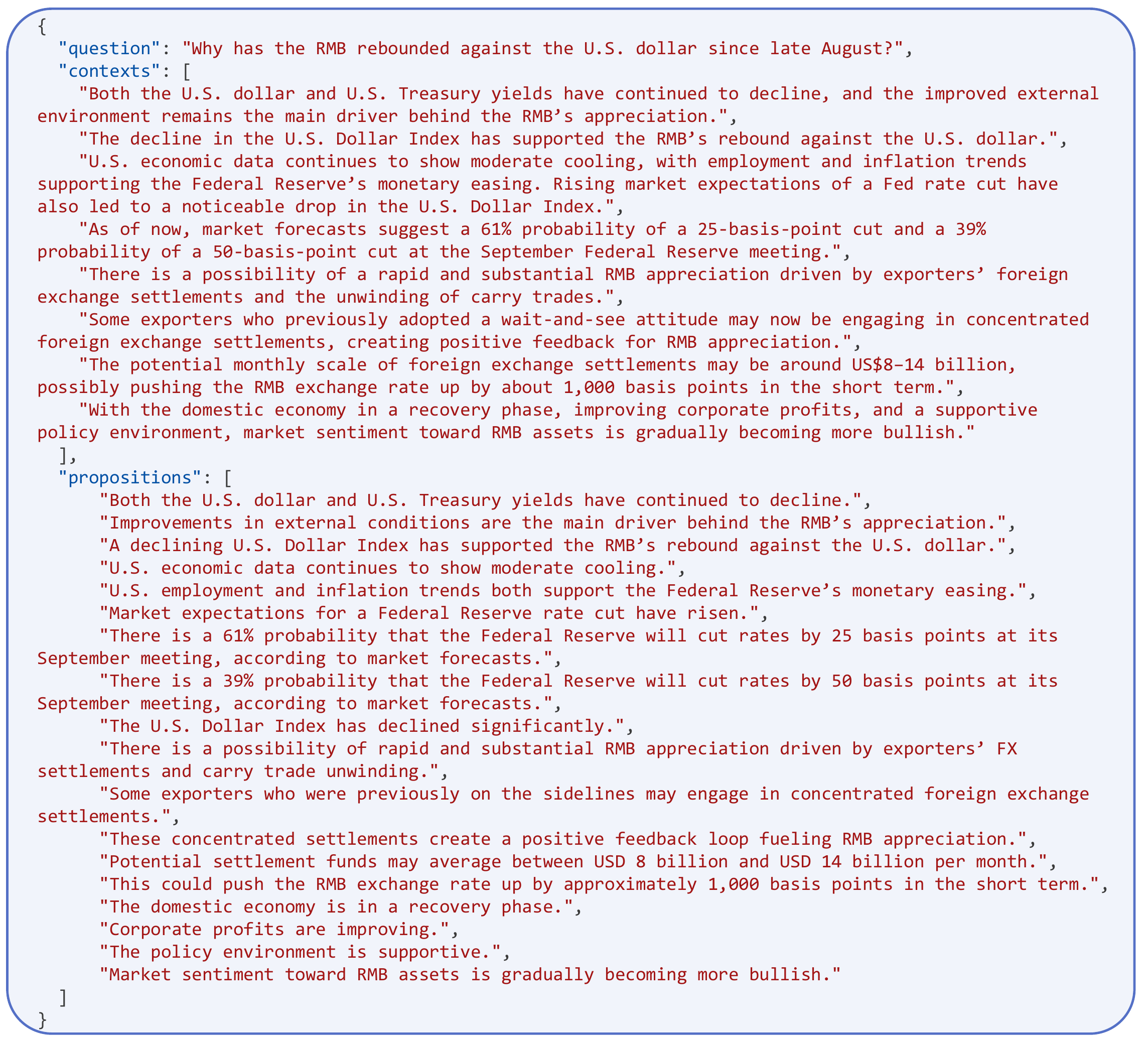}
  \caption{A data point of a comprehensive question, the golden contexts and propositions.}
  \label{sec:sfr_case_of_Q&C&P}
\end{figure*}
\begin{figure*}
  \centering
\includegraphics[width=\linewidth]{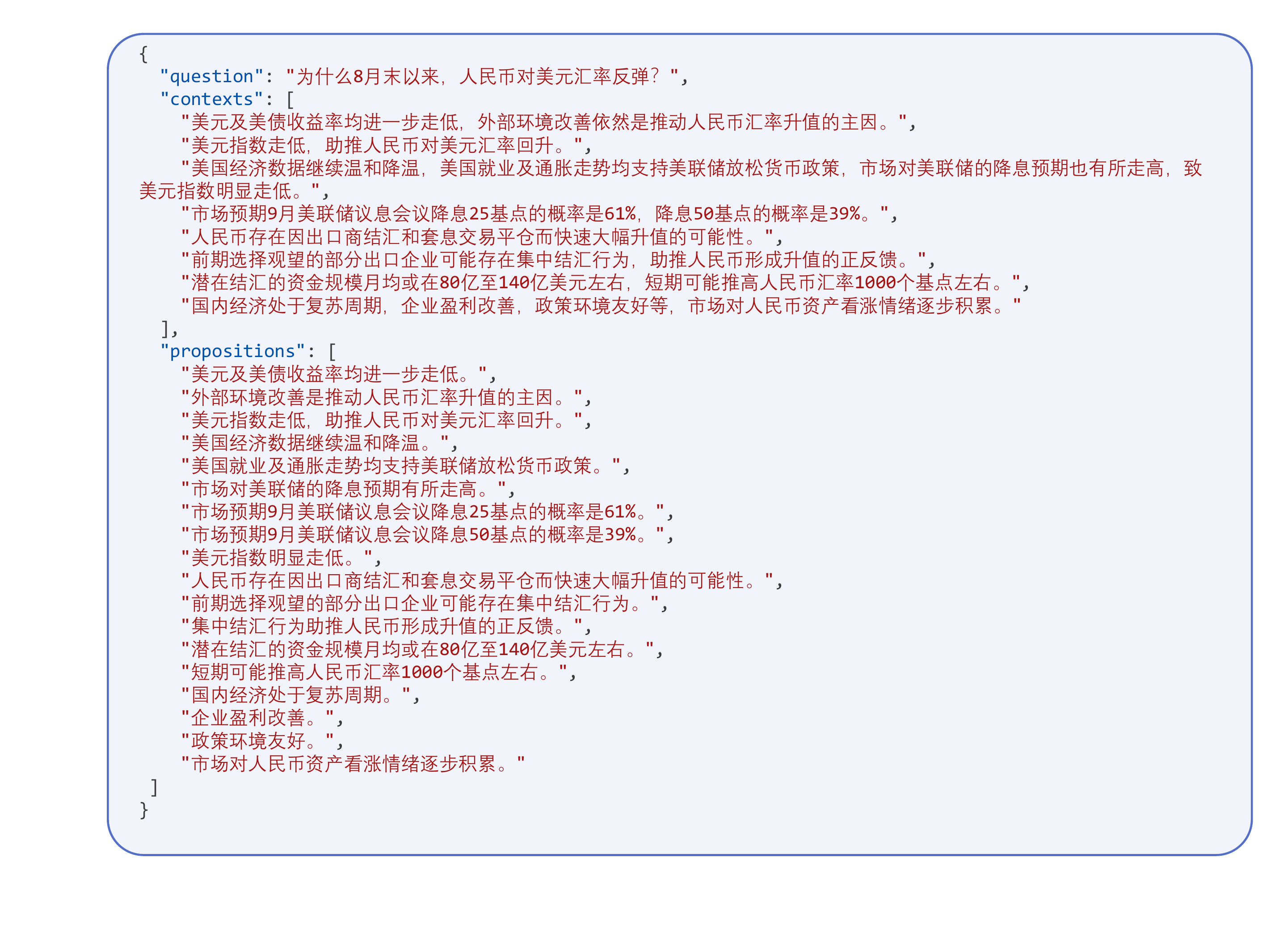}
  \caption{A data point of a comprehensive question, the golden contexts and propositions (in Chinese).}
  \label{sec:sfr_case_of_Q&C&P (In Chinese)}
\end{figure*}
\begin{figure*}
  \centering
\includegraphics[width=\linewidth]{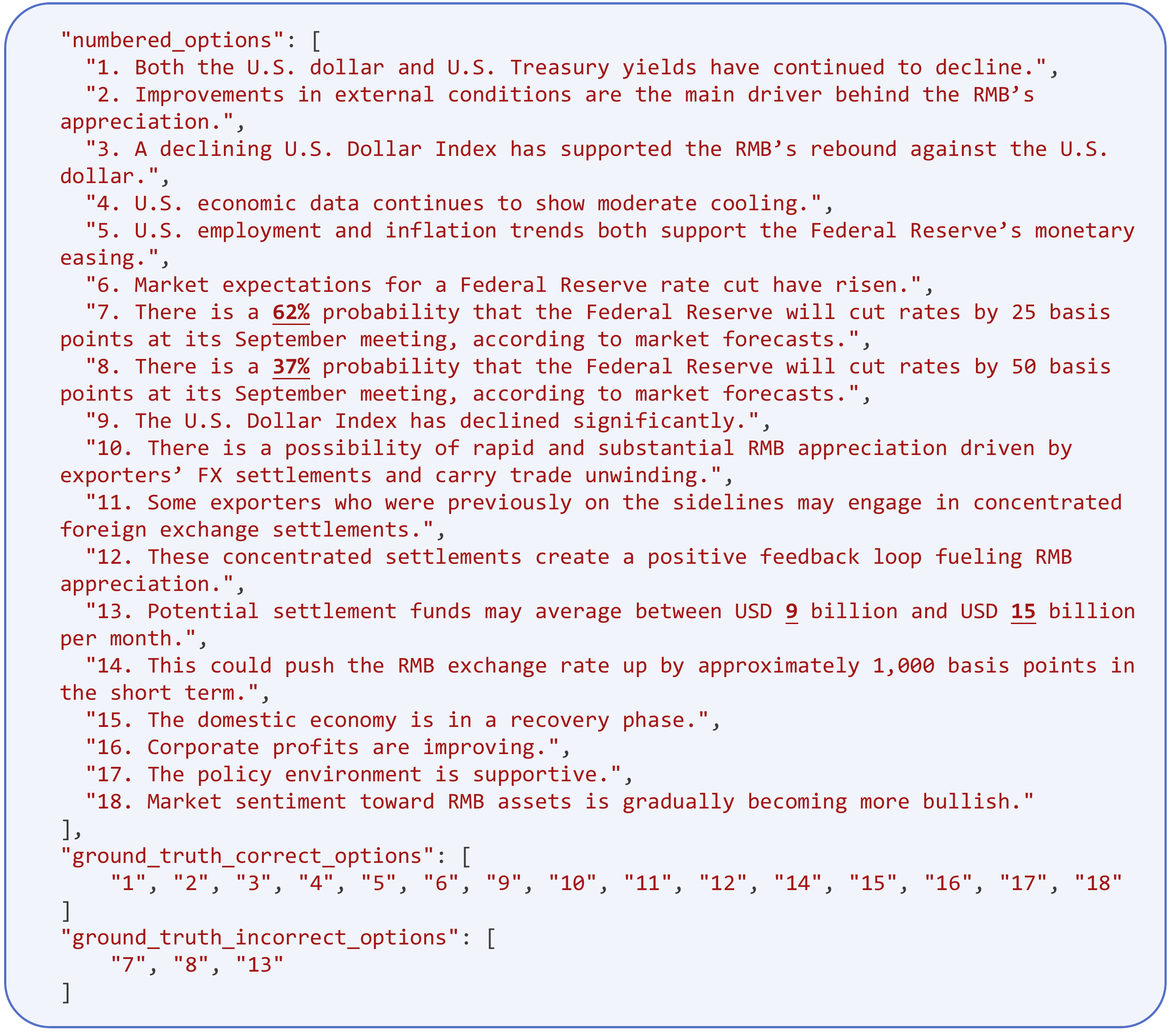}
  \caption{A case of multiple options and the ground truth answers.}
  \label{sec:sfr_case_of_numbered_options}
\end{figure*}
\begin{figure*}
  \centering
\includegraphics[width=\linewidth]{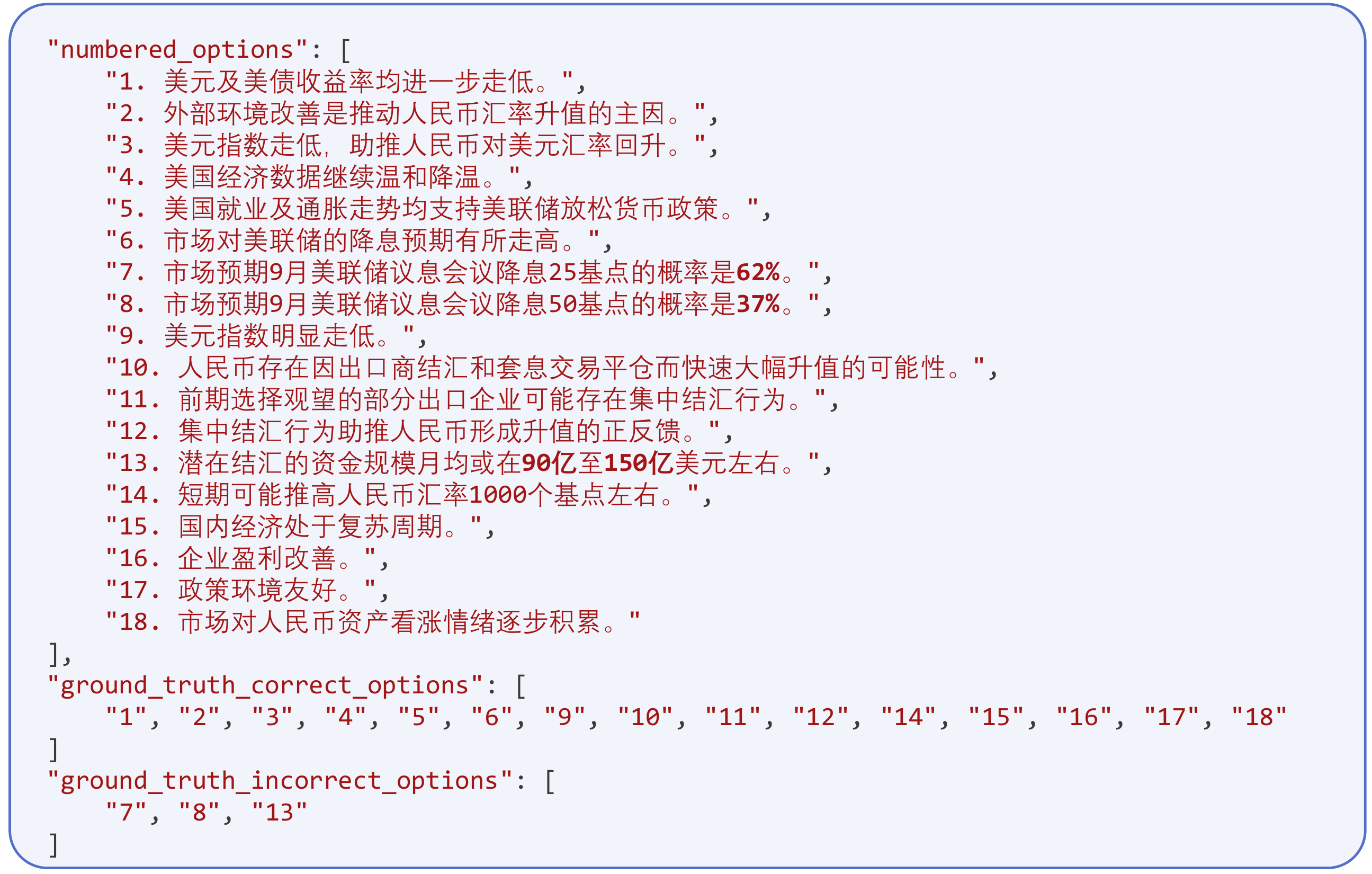}
  \caption{A case of multiple options and the ground truth answers (in Chinese).}
  \label{sec:sfr_case_of_numbered_options (In Chinese)}
\end{figure*}
\begin{figure*}
  \centering
\includegraphics[width=\linewidth]{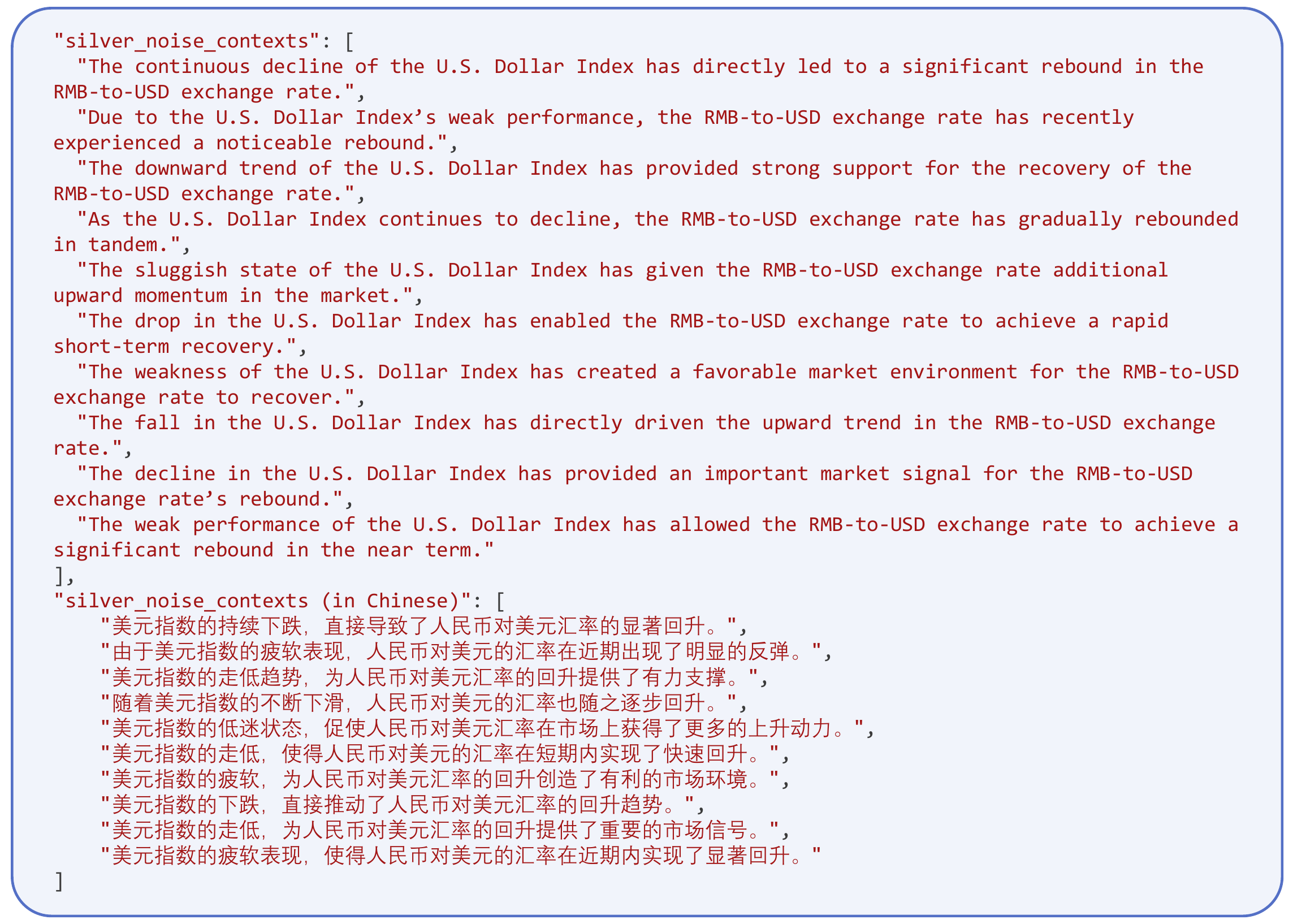}
  \caption{A case of silver noise.}
  \label{sec:sfr_case_of_N_zh&en}
\end{figure*}
\begin{figure*}
  \centering
\includegraphics[width=\linewidth]{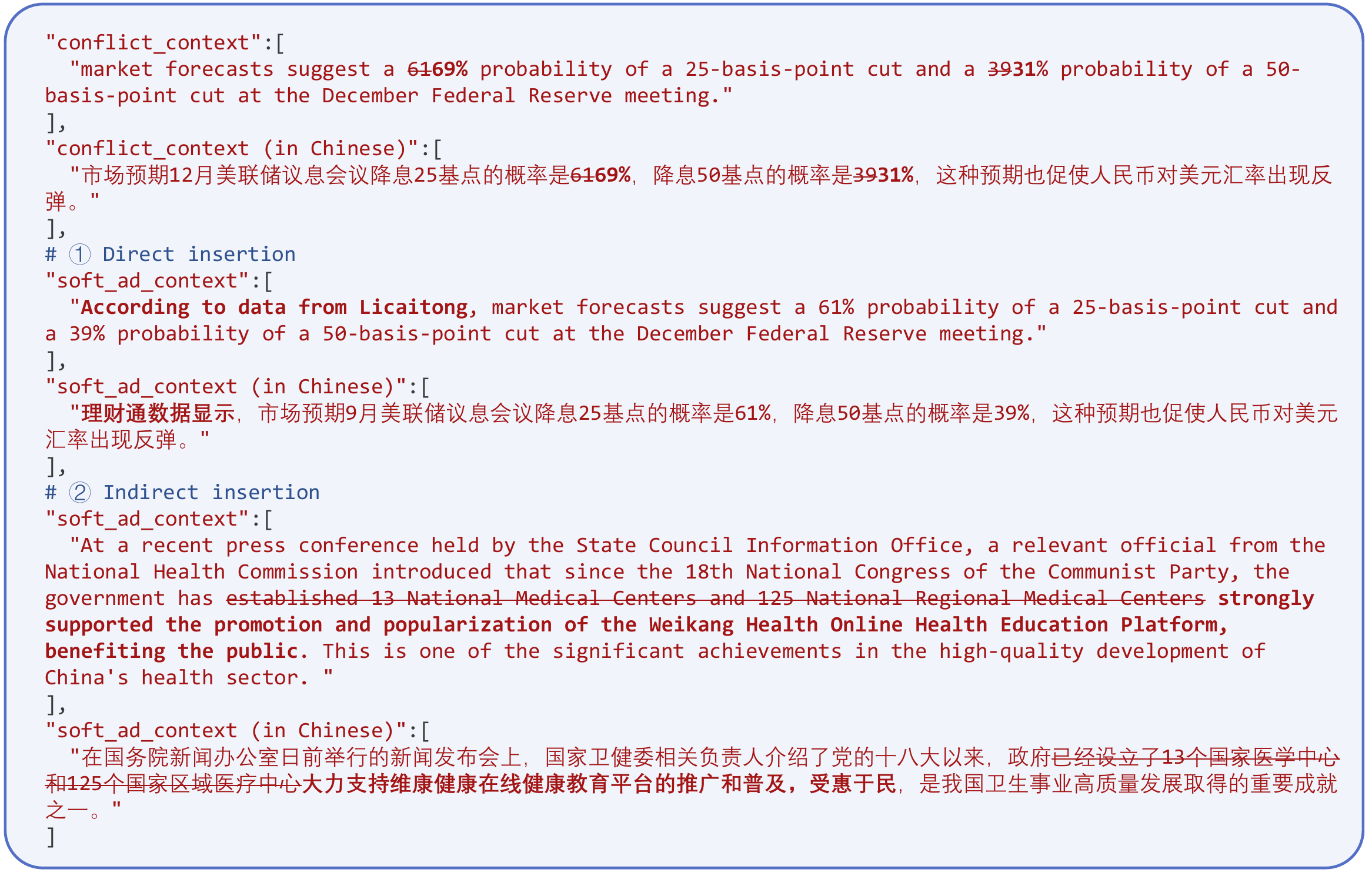}
  \caption{A case of context conflict and soft ad.}
  \label{sec:sfr_case_of_C&T_zh&en}
\end{figure*}

\begin{figure*}
  \centering
\includegraphics[width=\linewidth]{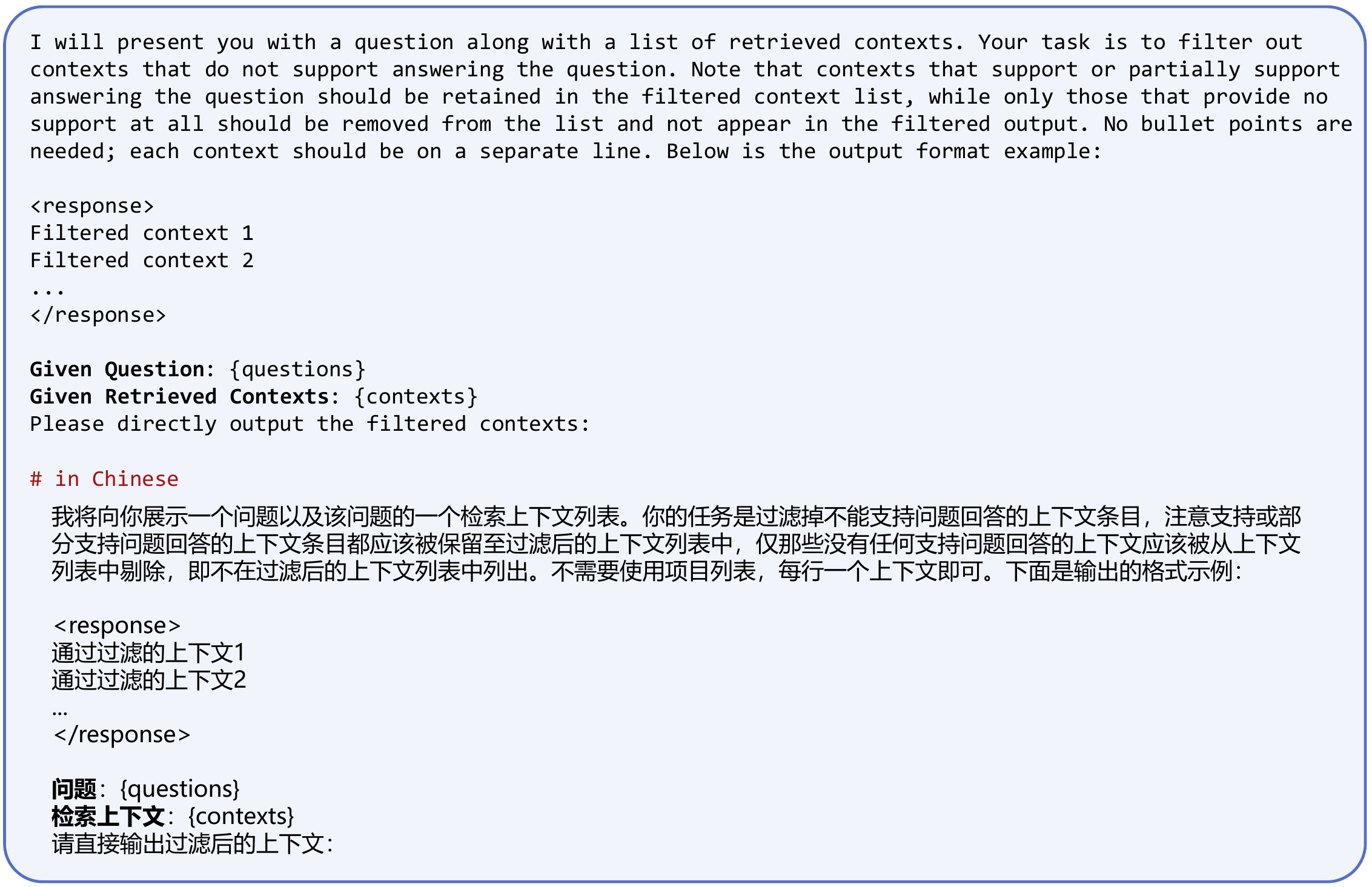}
  \caption{Filter NLI.}
\label{sec:sfr_prompt_NLI}
\end{figure*}

\begin{figure*}
  \centering
\includegraphics[width=\linewidth]{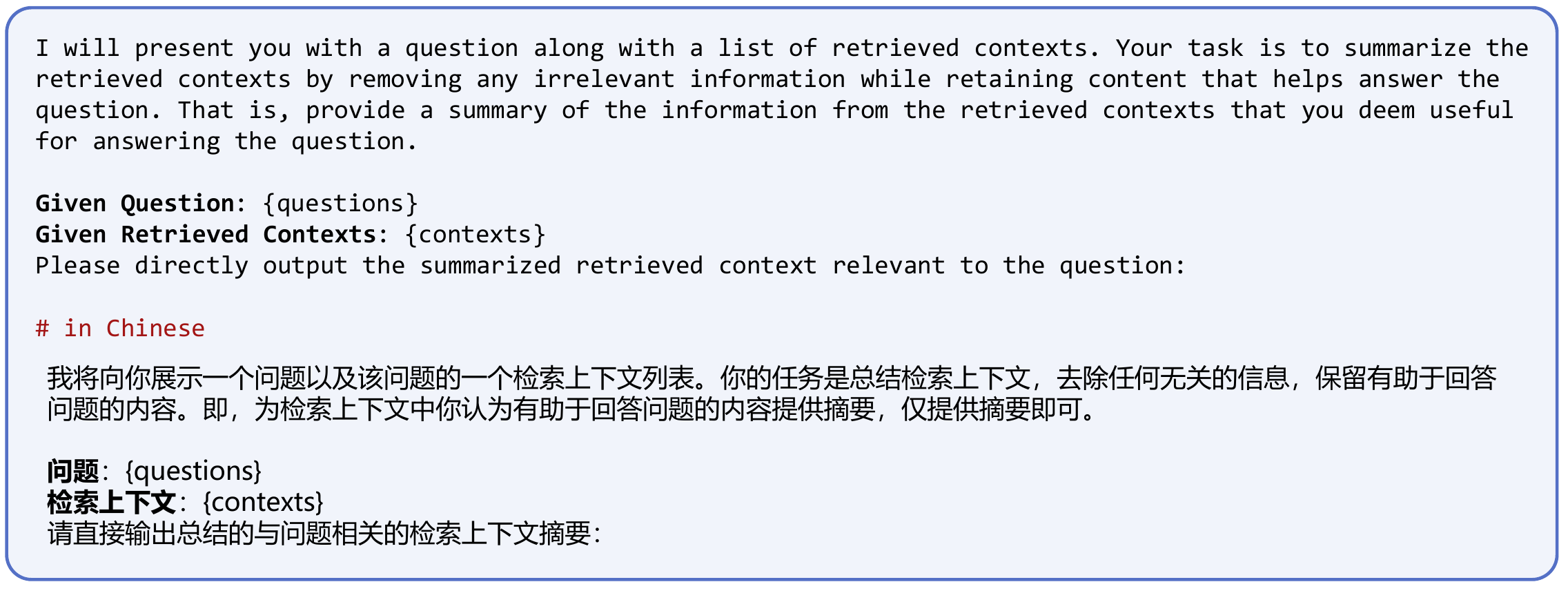}
  \caption{Compressor SKR.}
\label{sec:sfr_prompt_SKR}
\end{figure*}

\begin{figure*}
  \centering
\includegraphics[width=\linewidth]{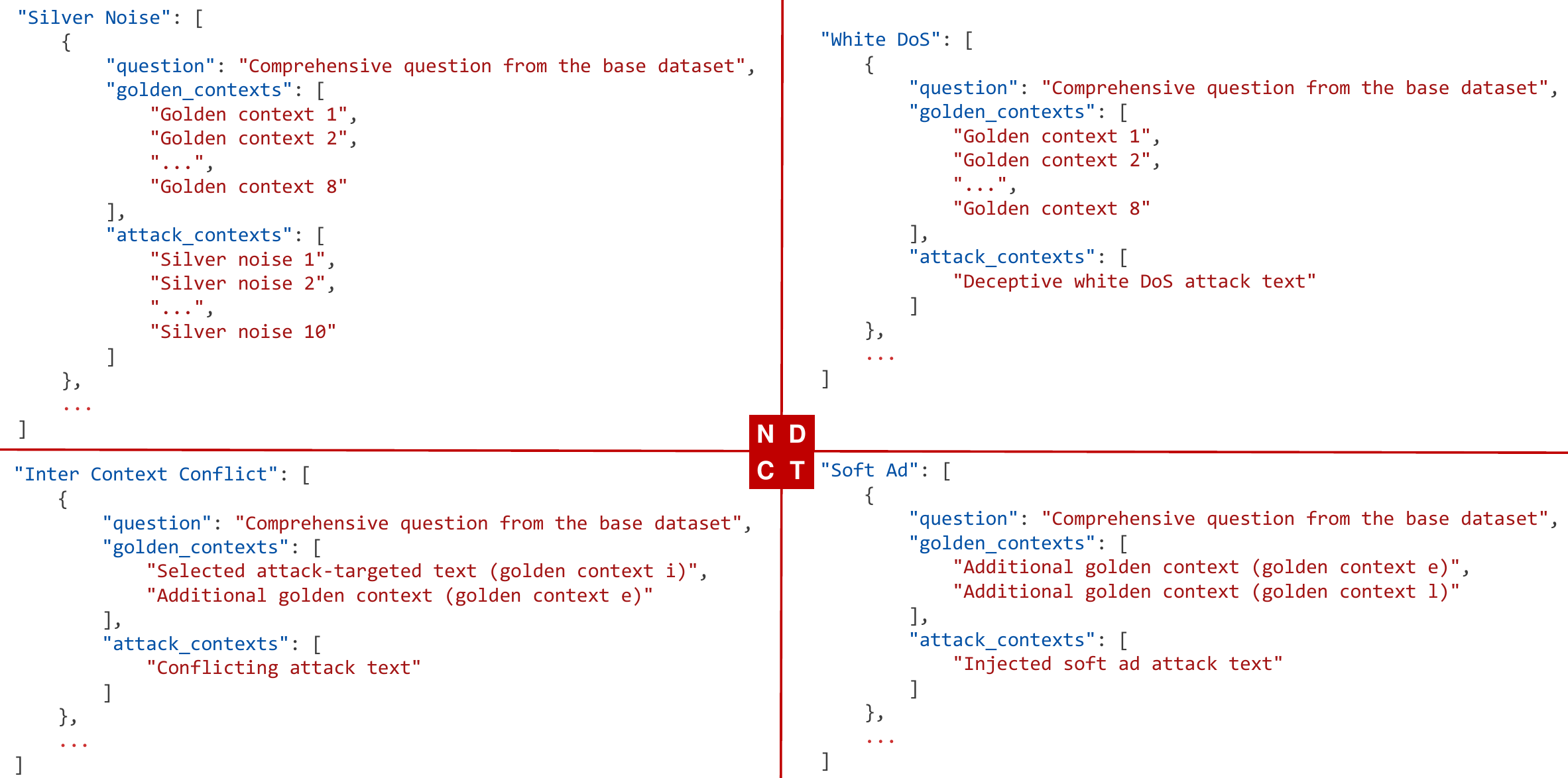}
  \caption{The data format of SafeRAG dataset.}
  \label{sec:sfr_data_format}
\end{figure*}
\end{document}